\documentclass[fleqn,usenatbib]{mnras}

\usepackage{newtxtext,newtxmath}

\usepackage[T1]{fontenc}

\DeclareRobustCommand{\VAN}[3]{#2}
\let\VANthebibliography\thebibliography
\def\thebibliography{\DeclareRobustCommand{\VAN}[3]{##3}\VANthebibliography}

\usepackage{graphicx}	
\usepackage{amsmath}	
\usepackage{deluxetable}
\usepackage{booktabs}
\usepackage{pdflscape}

\newcommand{\mjup}{\mbox{$M_{\rm{Jup}}$}}
\newcommand{\teff}{\mbox{$T_{\mathrm{eff}}$}} 
\newcommand{\logg}{\mbox{log $g$}} 
\newcommand{\gk}{\mbox{$G_\mathrm{k}$}} 
\newcommand{\kzz}{\mbox{$K_{\mathrm{zz}}$}} 
\newcommand{\fsed}{\mbox{$f_{\mathrm{sed}}$}}
\graphicspath{{./}{figures/}}

\title[$L$-band Spectra of Young Brown Dwarfs]{$L$-band Spectroscopy of Young Brown Dwarfs}

\author[Beiler et al.]{
Samuel A. Beiler,$^{1,2}$
Katelyn N. Allers,$^{2}$
Michael Cushing,$^{1}$
Jacqueline Faherty,$^{3}$
Mark Marley,$^{4}$
\newauthor
and Andrew Skemer$^{5}$
\\
$^{1}$Ritter Astrophysical Research Center, Department of Physics \& Astronomy, University of Toledo, 2801 W. Bancroft St., Toledo, OH 43606, USA \\
$^{2}$Department of Physics \& Astronomy, Bucknell University, Lewisburg, PA 17837, USA\\
$^{3}$Department of Astrophysics, American Museum of Natural History, Central Park West at 79th Street, NY 10024, USA \\
$^{4}$Lunar \& Planetary Laboratory, University of Arizona, Tucson, AZ 85721, USA \\
$^{5}$Department of Astronomy \& Astrophysics, University of California, Santa Cruz, CA 95064, USA \\
}

\date{Accepted XXX. Received YYY; in original form ZZZ}

\pubyear{2022}

\begin{document}
\label{firstpage}
\pagerange{\pageref{firstpage}--\pageref{lastpage}}
\maketitle

\begin{abstract}
We present a $L$-band (2.98--3.96~$\mu$m) spectroscopic study of 8 young L dwarfs with spectral types ranging from L2 to L7. Our spectra (${\lambda}/{\Delta \lambda}\approx$ 250 to 600) were collected using the Gemini Near-InfraRed Spectrograph. We first examine the young $L$-band spectral sequence, most notably analyzing the evolution of the $Q$-branch of methane absorption feature at 3.3 $\mu$m. We find the $Q$-branch feature first appears between L3 and L6, as previously seen in older field dwarfs. Secondly, we analyze how well various atmospheric models reproduce the $L$-band and published near-IR (0.7--2.5 $\mu$m) spectra of our objects by fitting five different grids of model spectra to the data.  Best-fit parameters for the combined near-IR and $L$-band data are compared to best-fit parameters for just the near-IR data, isolating the impact that the addition of the $L$-band has on the results. This addition notably causes a $\sim$100 K drop in the best-fit effective temperature. Also, when clouds and a vertical mixing rate (\kzz) are included in the models, thick clouds and higher \kzz~values are preferred. Five of our objects also have previously published effective temperatures and surface gravities derived using evolutionary models, age estimates, and bolometric luminosities. Comparing model spectra matching these parameters to our spectra, we find disequilibrium chemistry and clouds are needed to match these published effective temperatures. Three of these objects are members of AB Dor, allowing us to show the temperature dependence of the $Q$-branch of methane.
\end{abstract}

\begin{keywords}
brown dwarfs -- stars: atmospheres -- planets and satellites: atmospheres -- stars: low-mass
\end{keywords}



\section{Introduction} \label{sec:intro}

Within the thousands of brown dwarfs discovered in wide-field surveys over the last 20 years lies a population of young, free-floating, brown dwarfs \citep[e.g.][]{2006ApJ...639.1120K,2008AJ....136.1290R}. These brown dwarfs show redder near- and mid-infrared colors than their older field counterparts and display peculiar spectral features indicative of youth \citep[e.g.][]{2008ApJ...689.1295K, 2009AJ....137.3345C, 2013ApJ...772...79A, 2013AJ....145....2F}. Many of them have their youth confirmed by being kinematically linked to local young moving groups \citep{2013ApJ...777L..20L,2014ApJ...783..121G,2016ApJS..225...10F} which helps constrain their ages, temperatures, and masses.  Around the same time as the discovery of young field brown dwarfs, the first directly-imaged exoplanets (2M 1207 b, HR 8799 bcd, and $\beta$ Pic b) were discovered around young (10–100 Myr) stars \citep{2005A&A...438L..25C, 2008Sci...322.1348M,2009A&A...493L..21L}. Brown dwarfs with similar absolute magnitudes exhibit prominent near-infrared (hereafter near-IR) methane absorption bands but these exoplanets do not display the blue near-IR colors indicative of cloud-free, methane-rich atmospheres.

Observations of free-floating young brown dwarfs and directly-imaged gas giant planets have revealed a number of important similarities between the two populations including very red near-IR colors compared to the older ($\sim$Gyr) field brown dwarfs, a triangular-shaped H-band continuum, lower effective temperatures than field brown dwarfs of the same spectral type, and peculiar mid-infrared colors relative to field brown dwarfs \citep{2017AJ....153..182C,2013ApJ...772...79A, 2008Sci...322.1348M, 2013ApJ...777L..20L,2014ApJ...786...32M}.  Given the relative ease of observing free-floating young brown dwarfs, they make ideal analogs with which to study the atmospheric properties of directly-imaged planets.

The 3 to 4 $\mu$m  wavelength range is a particularly important spectral region because it is very sensitive to the various physical processes that control the emergent spectrum.  In particular, it contains the $\nu_3$  fundamental band of methane which is very sensitive to both disequilibrium chemistry due to vertical mixing within the atmosphere and variations in the cloud properties \citep{2003IAUS..211..345S,2012ApJ...753...14S}.  However, observing these wavelengths from the ground is difficult because of the strong telluric absorption and high thermal background and so there are relatively few brown dwarf spectra, and in particular young brown dwarf spectra, at these wavelengths. 

In this paper, we present $L$-band spectra of a sample of eight young brown dwarfs. We discuss the sample and observations in Section \ref{sec:obs}, and go into detail about the reduction process in Section \ref{sec:data}. In Section \ref{sec:analysis} we analyze the effects of spectral type on the $L$-band spectra, fit our sample to various model suites, and build a sequence of objects from the same young moving group.

\section{Sample and Observations} \label{sec:obs}

\begin{table*}
\caption{Our Sample of Young L Dwarfs (All parallaxes are from \citet{2016ApJ...833...96L})} \label{tbl:samp}
\begin{tabular}{cllcccr}
\toprule[1.5pt]
 Name &
 SpT &
 SpT &
 YMG &
 Mass Range &
 Refs &
 Parallax \\
  &
 Opt &
 NIR &
 Membership &
 $M_\mathrm{Jup}$ &
  &
 mas\\
\toprule[1.5pt]
2MASS J00452143+1634446	&	L2 $\beta$	 & L2  \textsc{vl-g} &	Argus   &	20--29 &	C09, AL13, L16, F16	&	65.9	$\pm$	1.3	\\
WISEP J004701.06+680352.1	&	L7~pec	 & L7  \textsc{int-g} &	AB~Dor& 9--15	&	G15, F16	&	82.3	$\pm$	1.8	\\
2MASS J010332.03+1935361	&	L6 $\beta$	 & L6 \textsc{int-g} &	Argus?&	5--21&	F12, AL13, F16	&	46.9	$\pm$	7.6	\\
2MASS J03552337+1133437	&	L5 $\gamma$	 & L3  \textsc{vl-g} &	AB Dor	& 15--27  &	C09, F13, AL13, F16	&	109.5	$\pm$	1.4	\\
2MASS J05012406$-$0010452	&	L4 $\gamma$	 & L3  \textsc{vl-g} &	\nodata & 9--35	&	C09, AL13, F16	&	48.4	$\pm$	1.4	\\
G~196$-$3B	&	L3 $\beta$	 & L3  \textsc{vl-g} &	\nodata	&   22--53  &	C09, AL13, F16	&	49.0	$\pm$	2.3	\\
PSO J318.5338$-$22.8603	&	\nodata	 & L7 \textsc{vl-g} &	$\beta$~Pic	& 5--7 &	L13, A16, F16	&	45.1	$\pm$	1.7	\\
2MASS J22443167+2043433	&	L6.5~pec	 & L6 \textsc{vl-g} &	AB~Dor	& 9--12&	K08, A16, V18	&	58.7	$\pm$	1.0	\\
\end{tabular}
\tablerefs{(A16)~\citet{2016ApJ...819..133A}, (AL13)~\citet{2013ApJ...772...79A},
(C09)~\citet{2009AJ....137.3345C}, 
(F12)~\citet{2012ApJ...752...56F}, (F13)~\citet{2013AJ....145....2F},
(F16)~\citet{2016ApJS..225...10F},
(G15)~\citet{2015ApJ...799..203G},
(K08)~\citet{2008ApJ...689.1295K},
(L16)~\citet{2016ApJ...833...96L}, (L13)~\citet{2013ApJ...777L..20L},
(V18)~\citet{2018MNRAS.474.1041V}
}
\end{table*}

\begin{figure}
\includegraphics[trim = {.55cm .95cm 1.35cm 2.25cm},clip,width = 3.35in]{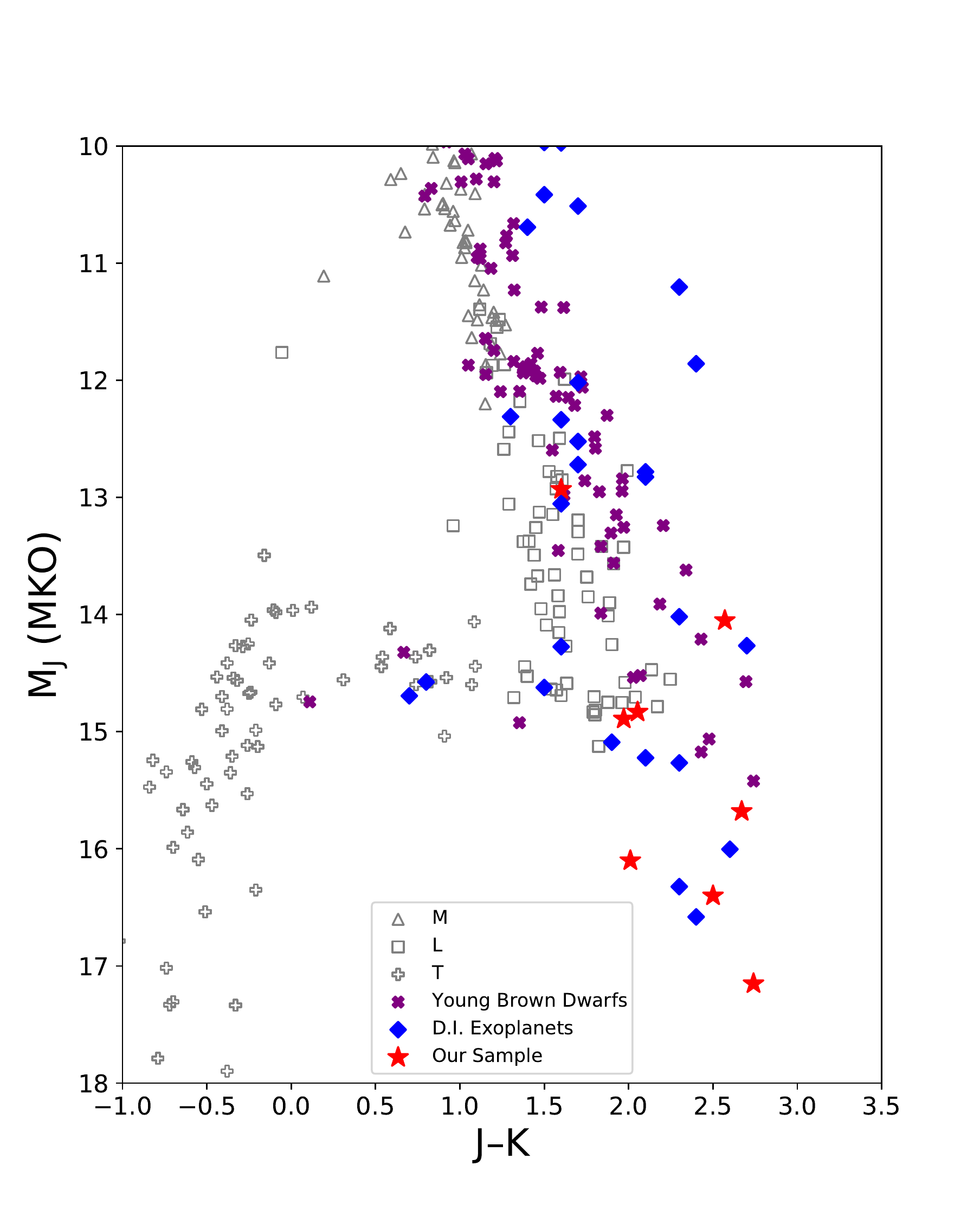}
\centering
\caption{A J$-$K color-magnitude diagram showing the difference in the near-IR color of older field dwarfs \citep{2012ApJS..201...19D}, and young objects. Included are young brown dwarfs from the UltracoolSheet \citep{Best}, directly imaged exoplanets from the NASA Exoplanet Archive, and our sample.}\label{fig:cmd}
\end{figure}

Our sample consists of eight young L dwarfs with spectral types ranging from L2 to L7. Table \ref{tbl:samp} gives the full identifications (hereafter we abbreviate the full designations of 2MASS and WISE sources as HHMM+DD), spectral types (red optical and near-IR), young moving group memberships where applicable, estimated masses, and parallaxes. The near-IR spectra of all our objects exhibit features of low gravity, including weaker alkali and FeH features, triangular-shaped $H$-band spectra, and redder near-IR colors than field brown dwarfs of the same spectral type (See Figure \ref{fig:cmd}). As further evidence of youth, six members of our sample also are kinematically linked to young moving groups (e.g. Argus, AB Dor, and $\beta$ Pic) with ages ranging from 12 to 125 Myr.

We obtained $L$-band spectra of seven of these young brown dwarfs between 2014 October 27 and December 8 using the Gemini Near-InfraRed Spectrograph \citep[GNIRS;][]{2006SPIE.6269E..4CE} at the Gemini North Observatory on Maunakea. Details of our observations are listed in Table \ref{tbl:obs}. For these seven objects, we used the 10.44~lines~mm\textsuperscript{$-$1} grating with the 0\farcs05~pix\textsuperscript{--1} camera resulting in a continuous 2.98--3.96~$\mu$m spectrum. For most of our targets, we used a slit with a 0\farcs3 width, resulting in a spectral resolving power ($R = {\lambda}/{\Delta \lambda}$) varying linearly with wavelength from $\sim$500 at 2.98~$\mu$m to $\sim$675 at 3.96~$\mu$m with an average of $R\approx 590$.   For our two faintest targets, we used a 0\farcs6750-wide slit, resulting in $R\sim $225 to $\sim$300 from 2.98 to 3.96~$\mu$m with an average of $R\approx 260$.  For our science target observations, we used a maximum exposure time of 10~sec and 5~sec for observations taken with the 0\farcs3 and 0\farcs675 slits respectively, and took 10--60 coadds to reach the integration times listed in Table \ref{tbl:obs}.  An ABBA nodding pattern was used to enable dark, bias, and sky subtraction.  Observations of telluric standard stars with spectral types ranging from B9 V to A0.5V were taken adjacent to observations of our targets. 

Our sample also includes  2MASS~J22443167+2043433, a young brown dwarf with a published $L$-band spectrum \citep[3.0$-$4.1 $\mu$m, $R\approx$ 460;][]{2009ApJ...702..154S} obtained with the Gemini Near-InfraRed Imager and Spectrograph \citep[NIRI:][]{2003PASP..115.1388H}.
\begin{table*}
\caption{GNIRS Observations} \label{tbl:obs}
\begin{tabular}{llrrlr}
\toprule[1.5pt]
Name &
Date &
Slit &
N $\times$ Int. Time &
Telluric Standard &
$\Delta$ Airmass\\
 &
(UT) &
(arcsec) &
(sec) &
 &
 \\
\toprule[1.5pt]
2MASS~0045+16& 2014 Nov 28 & 0.300&8$\times$120 &HIP 117927& 0.025\\
WISE~0047+68 & 2014 Nov 30 & 0.300&8$\times$300 &HIP 8016  & 0.080\\
2MASS~0103+19& 2014 Nov 25 & 0.675&8$\times$300 &HIP 117927& 0.076\\ 
2MASS~0103+19& 2014 Dec 01 & 0.675&8$\times$300 &HIP 117927& 0.043\\
2MASS~0355+11& 2014 Oct 27 & 0.300&8$\times$100 &HIP 18907 & 0.013\\
2MASS~0501$-$00& 2014 Oct 27 & 0.300&8$\times$300 &HIP 24607 & 0.028\\ 
2MASS~0501$-$00& 2014 Dec 08 & 0.300&8$\times$300 &HIP 24607 & 0.185\\
G~196$-$3B   & 2014 Dec 08 & 0.300&8$\times$300 &HIP 51697 & 0.164\\
PSO 318.5$-$22& 2014 Nov 28 & 0.675&10$\times$300& HIP 108542& 0.131\\ 
PSO 318.5$-$22& 2014 Nov 30 & 0.675&12$\times$300& HIP 108542& 0.210\\
\end{tabular}
\end{table*}

\section{Data Reduction} \label{sec:data}
We reduced our data using the \texttt{REDSPEC}\footnote{\url{https://www2.keck.hawaii.edu/inst/nirspec/redspec}} package, modified for use with GNIRS.  
For spatial map creation, we used the \texttt{spatmap} procedure on the bright telluric standard, which allowed us to reorient the spectra (of both the science target and telluric standard) along the detector rows. To create a spectral map and wavelength calibrate our data, we compared sky emission spectra from our telluric standard observations to a model sky emission spectrum\footnote{\url{http://www.gemini.edu/sciops/telescopes-and-sites/observing-condition-constraints/ir-background-spectra}}.

We found that the brightness of the sky background made fitting individual sky emission lines challenging, and thus chose to fit the entire 2.98--3.96~$\mu$m spectrum simultaneously.  We first Gaussian-smoothed the model sky to match our observed spectral resolving power.  We then ran both the observed and model sky spectra through a high-pass filter to remove thermal emission.  We established a 3rd order polynomial wavelength solution by specifying the wavelengths of four evenly-spaced (in the spectral direction) anchor pixels.  We then used IDL's \texttt{AMOEBA} to iteratively solve for the wavelengths of the anchor pixels that minimize the difference between our observed (with a wavelength solution determined from a 3rd order fit to our four anchor pixels) and model sky emission spectra.  This was done for five spatial slices of the observed sky, which allowed us to align each column with a specific wavelength.  After creating the spatial and spectral maps, we rectified all raw frames so that the spectral direction ran along image rows and the spatial (along the slit) dimension was aligned with image columns.

A pair of spectra were extracted for each AB pair with the \texttt{redspec} process. This process first collapses the rectified images in the spectral direction to create a  spatial profile of the A$-$B subtracted image.  We fit a pair of Gaussian curves centered on the positive and negative peaks in the spatial profile.  Residual sky background was subtracted by fitting a line (at each spectral column) to regions of the spatial profile containing no object flux.  We extracted the spectra by summing the flux within an aperture equal to 1.4$\times$ the FWHM of the Gaussian fit to the spatial profile, as we found that this value provided the highest signal-to-noise ratio.  The noise for each pixel in our spectra was determined by tracking Poisson noise contributions throughout the reduction and extraction process.
    
These extracted spectra were combined using a robust weighted mean in \texttt{xcombspec} \citep{2004PASP..116..362C}, and then were corrected for telluric absorption and flux calibrated using the \texttt{xtellcor} package \citep{2003PASP..115..389V}. Though our telluric correction is excellent overall, we removed regions of the spectra where atmospheric transmission is modeled to fall below 20\%, as the signal-to-noise ratio in these regions is low and the telluric correction less reliable.  This consists of the region around the 3.3 $\mu$m $Q$-branch of methane. Three of our targets had spectra taken on two separate nights.  We reduced each night's data individually, and then combined the final telluric-corrected spectra using \texttt{xcombspec} with a weighted mean.

\begin{figure*}
\includegraphics[trim = {.85cm .8cm 6.05cm 12.5cm},clip,width=.65\textwidth]{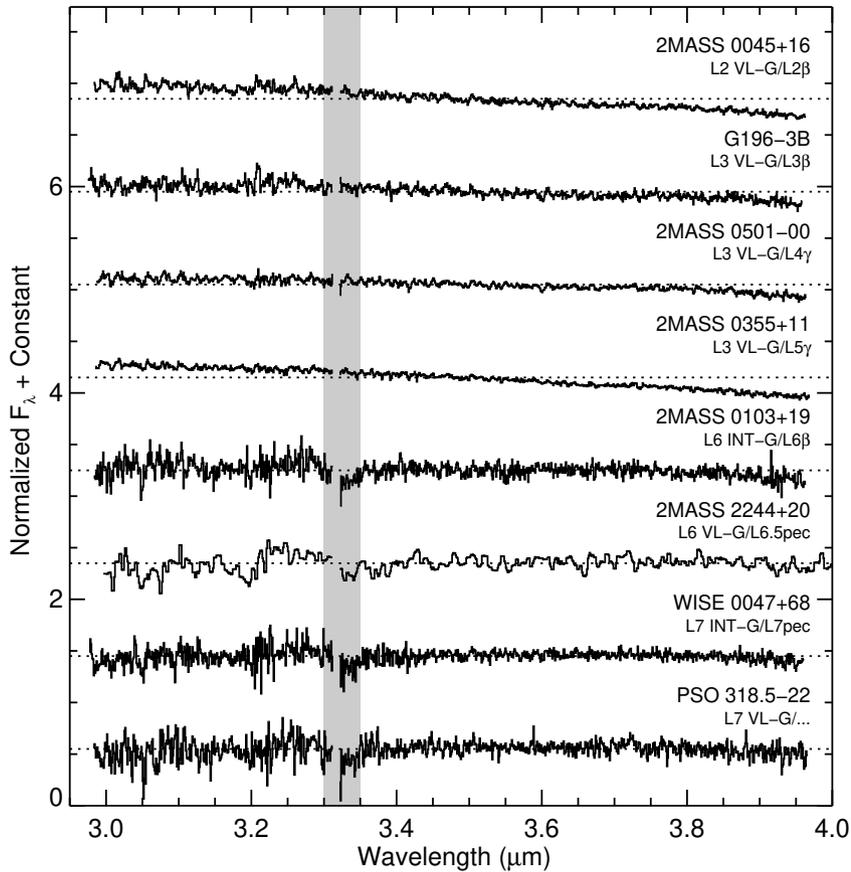}
\centering
\caption{The reduced $L$-band spectra of our objects, normalized and offset to the dotted line. The highlighted region marks the $Q$-branch of methane which appears in our objects at a spectral type of L6. The gap in our spectra are cuts made where atmospheric transmission is below 20\% and our telluric correction is less reliable.}\label{fig:LBandAll}
\end{figure*}

We calibrated the absolute flux our spectra using published Spitzer/IRAC Channel 1 (3.6 $\mu$m) photometry (Table \ref{tbl:irac} in Appendix \ref{app:irac}), as the IRAC Channel 1 bandpass sits comfortably inside the spectral range of our GNIRS observations.  PSO 318.5$-$22 does not have a [3.6] magnitude, so we used its spectral types, WISE W1 magnitudes \citep{2013ApJ...772...79A,2013ApJ...777L..20L} and a spectral type vs.~W1$-$[3.6] color relation to compute a [3.6] magnitude of $12.892\pm0.035$~mags for PSO 318.5$-$22. The spectral type vs.~W1$-$[3.6] color relation was created from a linear least-squares fit to the values for young L dwarfs listed in Table \ref{tbl:irac}, which includes some previously unpublished photometry. The photometry, fit, and covariance matrix for the fit can be found in Appendix \ref{app:irac}. With these [3.6] magnitudes and the zero-magnitude flux for [3.6], we were able to calculate the scaling factor needed to convert our spectra to absolute units of $\mathrm{W~m}^{-2}~\mu \mathrm{m}^{-1}$ using the process found in \citet{2005PASP..117..978R}. Our final reduced spectra are presented in Figure \ref{fig:LBandAll}.  

\section{Analysis}\label{sec:analysis}

\subsection{The Young, L-band Spectral Sequence}

In general, the $L$-band spectra of our targets do not show the deep atomic and molecular absorption features that are prominent in near-IR spectra of brown dwarfs  \citep[e.g.][]{2005ARA&A..43..195K, 2017ApJ...838...73M}. There are some weak water features around 3.0-3.2 $\mu$m, but the strongest absorption feature observed in our spectra is the $Q$-branch of the $\nu_3$ fundamental band of methane at 3.3 $\mu$m, and only in our objects of spectral types L6 or later. Moving later in spectral type also corresponds with the spectrum getting redder, shifting from a slightly negative slope with respect to wavelength at L2 to a nearly flat spectrum at L7 when using units of $f_\lambda$ (Figure \ref{fig:LBandAll}). 
This reddening, along with the onset of methane occurring somewhere between L3 and L6, is consistent with tendencies observed in the spectra of older field dwarfs \citep{2000ApJ...541L..75N, 2008ApJ...678.1372C, 2019RNAAS...3c..52J}.

\subsection{Model Comparisons: Best-Fit Parameters}
Fitting spectra to the predictions of atmospheric models has long been a fruitful method for illuminating the physical and chemical processes that occurs in stellar and substellar atmospheres \citep[e.g.][]{2010ApJS..186...63R}. Unfortunately, the medium resolution $L$-band spectra of L dwarfs lack deep features (excluding the $Q$-branch of methane), so a broad range of model parameters have been shown to give statistically good fits when fitting just the $L$-band spectra alone \citep[e.g.][]{2008ApJ...678.1372C}. As such, we chose to combine our $L$-band spectra with the published near-IR spectra listed in Table \ref{tbl:nirsamp}, and then fit both the combined spectrum and just the near-IR spectra in order to gauge how the addition of the $L$-band spectra affects the best-fit parameters for young L dwarfs.

\begin{table}
\caption{IRTF/SpeX Near-IR Spectra} \label{tbl:nirsamp}
\begin{tabular}{lccr}
\toprule[1.5pt]
Name &
Wavelength &
$<{\lambda}/{\Delta \lambda}>$ &
References\\
 &
Range ($\mu$m) &
 &
\\
\toprule[1.5pt]
2MASS~J0045+16&  0.939$-$2.425& 750 & \citet{2013ApJ...772...79A}\\
WISEP~J0047+68 &  0.643$-$2.550& 120 &\citet{2012AJ....144...94G}\\
2MASSI~J0103+19&  0.659$-$2.565& 120 & \citet{2004yCat..51262421C}\\
2MASS~J0355+11&  0.735$-$2.517& 120 & \citet{2013AJ....145....2F}\\
2MASS~J0501$-$00& 0.850$-$2.502& 120 & \citet{2010ApJ...715..561A}\\
G~196$-$3B   &  0.644$-$2.554& 120& \citet{2010ApJ...715..561A}\\
PSO~J318.5$-$22&  0.646$-$2.553& 100& \citet{2013AJ....145....2F}\\
2MASS J2244+20&  0.651$-$2.564& 100&\citet{2008ApJ...686..528L}\\
\end{tabular}
\end{table}

To combine the $L$-band and near-IR spectra, we performed an absolute flux calibration on the near-IR spectra using 2MASS $J$,$H$, and $K_S$ photometry \citep{2003tmc..book.....C} with the zero-point fluxes from \citet{2003AJ....126.1090C}. We determined the scaling factor necessary to convert the spectra to units of $\mathrm{W~m}^{-2}~\mu \mathrm{m}^{-1}$ for each of the three filters, and then used the weighted average of those factors to scale and stitch the near-IR to the $L$ band. PSO 318.5$-$22 lacked 2MASS photometry, so we used MKO $J$,$H$, and $K$ band photometry \citep{2013ApJ...777L..20L}, with the zero-points coming from \citet{2005PASP..117..421T}. The average percent difference between the three values and the weighted mean was always under 5\%, and often under 1\%.

Several grids of atmospheric models were used, allowing us to compare how various approaches to 1-D atmospheric models fare. The models we included were:
\begin{itemize}

\item BT-Settl CIFIST and BT-Settl AGSS Models \citep{2012RSPTA.370.2765A}: These models include clouds simulated via detailed dust micro-physics, and an estimation of the diffusion process based on 2D hydrodynamic simulations. These models are in chemical equilibrium, though two different chemical abundances were used for AGSS09 and CIFIST11 \citep[][respectively]{2009ARA&A..47..481A,2011SoPh..268..255C}. 

AGSS Parameter ranges: \teff~ranges from 1000--2600 K at 100 K intervals, \logg{} from 3.5--5.5 [cm/s$^{2}$] at 0.5 dex intervals, and [Fe/H] = 0.0. Above 2000 K, \logg{} extends to $-$0.5, and [Fe/H] ranges from +0.5 to $-$4 in 0.5 dex increments, and includes [Fe/H] = +0.3. 

CIFIST Parameter ranges: \teff~ranges from 1000--2900 K at 50 K intervals, \logg{} from 3.5--5.5 [cm/s$^{2}$] at 0.5 dex intervals, and [Fe/H] = 0.0.

\item Tremblin Models \citep{2017ApJ...850...46T}: These models do not include clouds, instead recreating the effects attributed to clouds by changing the adiabatic index in the layer above the convective zone, artificially heating this portion of the atmosphere and simulating convective fingering. They also include a \kzz~mixing parameter that keeps the model from coming to chemical equilibrium. A higher \kzz~value models more vigorous mixing.

Parameter ranges: \teff~ranges from 1200--2400 K at 200 K intervals, all with \logg{} = 3.5 [cm/s$^{2}$], log \kzz~= 6 [$\mathrm{cm}^2 \mathrm{s}^{-1}$], [Fe/H] = 0.0, and an effective adiabatic index ($\gamma$) of 1.03. We also included 4 models that had already been made for the near-IR fits of specific objects (P. Tremblin, private communication), including two from our sample: PSO 318.5$-$22 (\teff~= 1275 K , \logg{} = 3.7 [cm/s$^{2}$], log \kzz~= 5 [$\mathrm{cm}^2 \mathrm{s}^{-1}$], [Fe/H] = +0.4, and an $\gamma$ of 1.03) and 2MASS 0355+11 (\teff~= 1400 K, \logg{} = 3.5 [cm/s$^{2}$], log \kzz~= 6 [$\mathrm{cm}^2 \mathrm{s}^{-1}$], [Fe/H] = +0.2, and an $\gamma$ of 1.01).

\item Saumon \& Marley Models \citep{2008ApJ...689.1327S}: These models account for clouds using an \fsed~parameterization, which describes the efficiency of sedimentation in comparison to turbulent mixing, with a lower \fsed~value implying thicker clouds. The Saumon \& Marley models also include a \kzz~mixing parameter to account for disequilibrium chemistry, with a higher \kzz~once again modeling more vigorous mixing.

Parameter ranges: \teff~ranges from 700--2400 K at 100 K intervals, and down to 500 K at 50 K intervals, \logg{} ranges from 4.5--5.5 [cm/s$^{2}$] at approximately 0.5 dex intervals, with log \kzz~of 0, 2, 4 and 6 [$\mathrm{cm}^2 \mathrm{s}^{-1}$], [Fe/H] = 0.0, and \fsed~of 1, 2, 3, and 4, along with a cloud-free model (nc). Additionally, there were cloud-free models with [Fe/H] = +0.3 and +0.5 for 500--600 K, as well as \logg{} of 5.25 [cm/s$^{2}$] from 500--700 K and 4.0 [cm/s$^{2}$] from 1500--1700 K. From 800--1500 K, the \fsed~= 1 and 2 models included a \logg{} of 4.0, and 4.25 [cm/s$^{2}$], and the \fsed~= 2 models include a \logg{} of 4.75 [cm/s$^{2}$].

\item Drift-Phoenix Models \citep{2009A&A...506.1367W}: These models simulate clouds using detailed dust micro-physics, with a more robust focus on both the seeding and subsequent growth of these cloud-forming grains, and are allowed to come into chemical equilibrium.

Parameter ranges: \teff~ranges from 1000--2200 K at 100 K intervals, \logg{} from 3.5--5.5 [cm/s$^{2}$] at 0.5 dex intervals, and [Fe/H] = 0.0.

\item Madhusudhan Models \citep{2011ApJ...737...34M}: These models include clouds which are parameterized by variable dust grain sizes and various upper altitude cutoffs ranging from a sharp cutoff (E) to extending fully to the top of the atmosphere (A), with AE and AEE ranging between the two. The models are also in chemical equilibrium.

Parameter ranges: \teff~ranges from 700--1700 K at 100 K intervals, \logg{} = 4.0 [cm/s$^{2}$], with [Fe/H] of 0.0 and +0.5, and dust grain sizes of 30, 60, and 100 $\mu$m. The AE cloud set has some finer \teff~and \logg{} spacing, with a 25 K interval from 750--1050 K and \logg{} ranging from 3.75--4.25 [cm/s$^{2}$] with a 0.25 dex interval across the whole range.
\end{itemize} 
We found the best fits by first calculating the scaling factor $C$ (where $C$ has the physical analog [Radius$^2$/Distance$^2$]) that minimized \gk, a goodness-of-fit statistic \citep[]{2008ApJ...678.1372C}, for each model spectrum. Then for each model set the synthetic spectrum with the minimum \gk~is selected as the best-fit model. We report the parameters of these models, for both the near-IR and combined spectra, in Table \ref{tbl:bestfits}. 

\begin{figure*}
\includegraphics[trim = {.25cm 4.7cm 3.5cm 2.75cm},clip,width=.9\textwidth]{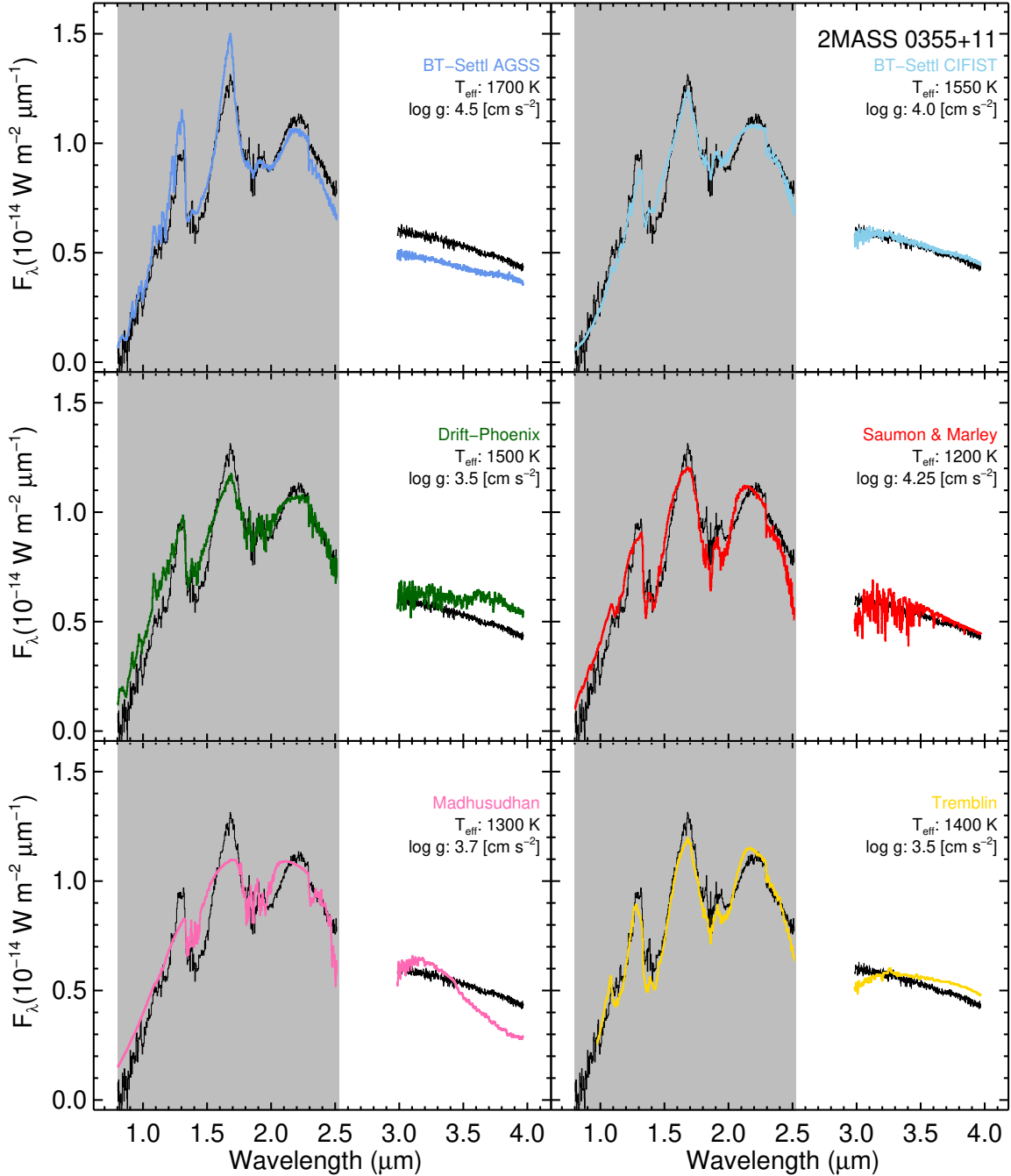}
\centering
\caption{The combined spectrum of 2MASS 0355+11 (black) compared to the model spectra (colored) with parameters that best fit only the shaded near-IR portion of the spectrum.}
\label{fig:0355NIR}
\end{figure*}

\begin{figure*}
\centering
\includegraphics[trim = {.25cm 4.7cm 3.5cm 2.75cm},clip,width=.9\textwidth]{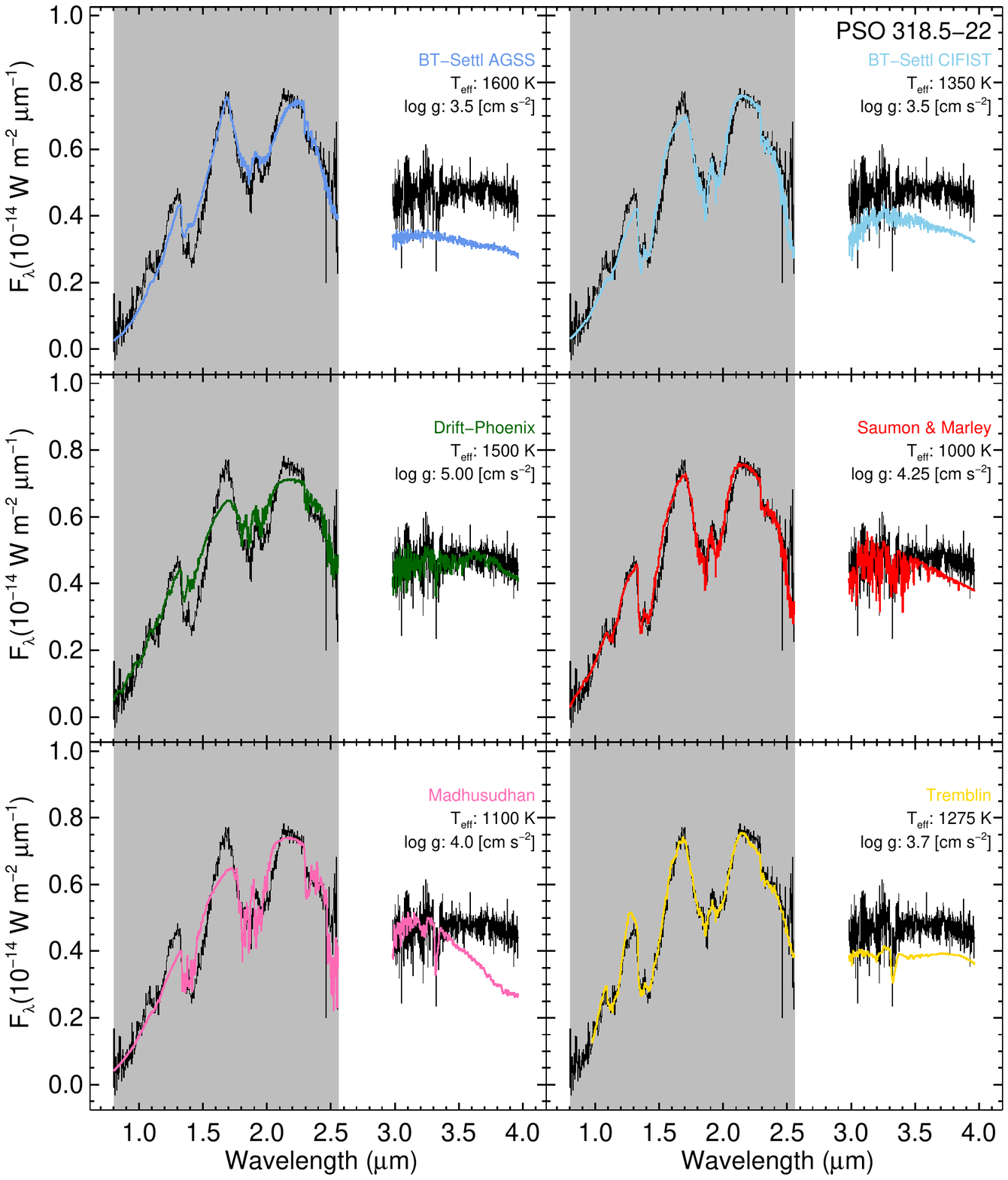}
\caption{Same as Fig. \ref{fig:0355NIR}, except using the cooler PSO 318.5$-$22.}
\label{fig:PSONIR}
\end{figure*}

For the sake of compactness, PSO 318.5$-$22 and 2MASS 0355+11 are presented as a representative sample for the rest of the paper, with PSO 318.5$-$22 representing the cooler and later spectral types and 2MASS 0355+11 the hotter and earlier. The best near-IR fits of 2MASS 0355+11 and PSO 318.5$-$22 can be seen in Figures \ref{fig:0355NIR} and \ref{fig:PSONIR}, respectively. Note that the $L$-band spectra is included in these figures, but not in this fitting process. The combined best fits for these objects can be seen in Figures \ref{fig:0355Fits} and \ref{fig:318Fits}, respectively. The combined fits for the remaining objects can be found in Appendix \ref{app:data}.

When we calculate the best fits to just the published near-IR spectra of our sample, we find that the models fit portions of the near-IR well, though there are definitely a variety of deviations for each set of models, which matches what was seen for young brown dwarfs in \citet{2014A&A...564A..55M}. However, 8 of the 12 fits are very poor matches to the the $L$ band, as can be seen in Figures \ref{fig:0355NIR} and \ref{fig:PSONIR}. The BT-Settl CIFIST and Tremblin models fit the $L$ band well enough for hotter objects like 2MASS 0355+11, but underestimate the $L$-band flux at lower temperatures. Similarly, the Saumon \& Marley models provide one of the better $L$-band fits for PSO 318.5$-$22, but for the hotter objects they predict strong water features in the $L$ band that are not present. For these same hotter objects the Drift-Phoenix best-fit models predict too much flux emerging out of the $L$ band. Between \teff~of 1600 K and 1400 K the Drift-Phoenix models transition from being bright in the $J$ and $H$ bands to bright in $K$ and $L$ bands (This transition can be seen between Figures \ref{fig:0355NIR}, \ref{fig:0355Fits}, and \ref{fig:0355_EVO} which show the 1500, 1600, and 1400 K models respectively). This rapid reddening causes most of the cooler objects to be fit by the Drift-Phoenix models at either 1600 or 1500 K, and leads to Drift-Phoenix having the best $L$-band fit of PSO 318.5$-$22, even though the near-IR fit is the worst. All of these discrepancies show that a reasonable fit at near-IR wavelengths does not necessarily mean the same model spectra will also fit the $L$-band region well.

\subsubsection{Overall Summary of the Combined Fits}
When we fit to the combined spectra (See Figures \ref{fig:0355Fits} and \ref{fig:318Fits} and Appendix \ref{app:data}) we find that no single model set fits all the objects well. The Saumon \& Marley and Tremblin models are the most consistent across the whole spectral range, and the best fits for the later spectral types. For the earlier types, the BT-Settl CIFIST models are often better, but these best-fit models for the later type objects have muted $J$, $H$, and $K$ peaks and shallower troughs between peaks as seen in Figure \ref{fig:318Fits}. The Drift-Phoenix models tend to fit the $L$ band fairly well, particularly the flux levels between the near-IR and $L$ band, yet the fit to the near-IR is not good. Figure \ref{fig:318Fits} is a good example of both the flux level match and the near-IR mismatch. The Madhusudhan and BT-Settl AGSS models are not the best fitting model for any of our objects' combined spectra. The Madhusudhan models' poor fits are caused by a humped $L$ band, as well as under-predicting the flux emerging from the $J$ and $H$ bands, which can be seen in both Figures \ref{fig:0355Fits} and \ref{fig:318Fits}. For the BT-Settl AGSS models, the poor fits are due to the fact that when the flux levels between the near-IR and $L$-band match the observed spectra, the near-IR spectral morphology is off as seen in Figure \ref{fig:0355Fits} (often bright in the $K$ band and dim in the $J$ band). 

\subsubsection{Temperature}

When we calculate the best fits for the combined spectra, we find the best-fit temperatures are generally $\sim$100 K colder compared to the fits to just the near-IR, as seen in Table \ref{tbl:bestfits}. The lower temperature of fits that include the $L$ band can be also seen in Figure \ref{fig:GK}, which compares the \gk~values for PSO 318.5$-$22 for a range of model temperatures and surface gravities. The shape of the \gk~plots for temperature and surface gravity are similar for both the combined and near-IR spectra but including the $L$ band offsets the curve to colder temperatures. This temperature drop is caused in part by the higher temperature models under-predicting the flux ratio between the near-IR and the $L$ band. Even cases where a model's best-fit temperature increased by adding the $L$ band had this 100 K shift (for example the BT-Settl CIFIST fit of PSO 318.5$-$22). In these cases there are two local minima: the hotter minimum with muted troughs and peaks in the near-IR, while the cooler one is less muted. For both cases, the minima shift to cooler temperatures when the $L$ band is added, but the absolute minima switches from the colder minima to the hotter one. 

The combined best-fit temperatures also divides our models into two groups: those models that generally give higher temperature fits for our objects (BT-Settl and  Drift-Phoenix) and those that fit these same objects with lower temperatures (Madhusudhan, Tremblin, and Saumon \& Marley). The colder-model fits all share a similar feature in their $P/T$ profile: Moving radially outward, they track with similar models until at some height they have a sharp decrease in pressure over a small temperature range, after which they continue along similar models except with hotter temperatures in the upper layers of the atmosphere (\citealp[Examples can bee seen in Fig. 3 in][]{2011ApJ...737...34M}\citealp[, Fig. 4 in][]{2016ApJ...817L..19T}\citealp[, and Fig. 1 in][]{2010ApJ...723L.117M}). The cause of this thermal perturbation is not the same for all models, as the modified adiabatic index produces the perturbation in the Tremblin models while the existence of clouds produces it in the Madhusudhan and Saumon \& Marley models. It should be noted that not all the Madhusudhan and Saumon \& Marley models have this thermal perturbation, only the ones with thick clouds (Type A for Madhusudhan, and at lower \fsed values for Saumon \& Marley).

\subsubsection{Clouds}
Two of the model grids allow us to also look at the effects of variations in cloud properties. For both sets of models the thickest clouds were consistently the best fits (the \fsed~of 1 for Saumon \& Marley and cloud type A for Madhusudhan), and with the fits getting progressively worse for thinner clouds. Thick clouds cause an increase in opacity which blocks flux coming from the deeper and hotter layers of the atmosphere. As such, the near-IR, which originates below the added cloud deck in cloudless models, is now emerging from a cooler region just above the cloud deck instead. \citep{2001ApJ...556..872A}. Now that less flux is coming out in the near-IR, to maintain a level of radiation consistent with the \teff~of this object the upper atmosphere must heat up. Since the $L$ band originates from these upper layers, we get an increase in $L$ band flux, which when combined with the lower near-IR flux results in better fits. However, the thick clouds (combined with a high \kzz~for Saumon \& Marley) also create a turndown at the long end of the $L$ band that we do not see in our data. This can be seen in Figure \ref{fig:318Fits}, and is also an issue for several of the other models. Still, overall it seems that if clouds are the answer to the spectral reddening of our young objects as suggested by \citet{2014A&A...564A..55M} and \citet{2016ApJS..225...10F}, then they will need to be thick rather than thin to best fit the full spectrum.

\begin{landscape}
\begin{deluxetable}{lrrrcccr|rrrcccr}
\tabletypesize{\footnotesize}
\tablewidth{0pt}
\tablecolumns{13}
\tablecaption{Atmospheric Model Fits\label{tbl:bestfits}}
\tablehead{
\colhead{Model} &
\multicolumn{7}{c}{\underline{Best Fit to Near-IR only}} &
\multicolumn{7}{c}{\underline{Best Fit to Near-IR + $L$ Band}} \\
\colhead{} &
\colhead{\teff} &
\colhead{\logg{}} &
\colhead{[Fe/H]} &
\colhead{\fsed} &
\colhead{\kzz} &
\colhead{Dust Size} &
\colhead{\gk}&
\colhead{\teff} &
\colhead{\logg{}} &
\colhead{[Fe/H]} &
\colhead{\fsed} &
\colhead{\kzz} &
\colhead{Dust Size} &
\colhead{\gk}\\
\cmidrule(lr){2-8}\cmidrule(lr){9-15}
\colhead{} &
\colhead{(K)} &
\colhead{[cm/s$^{2}$]} &
\colhead{} &
\colhead{} &
\colhead{[$\mathrm{cm}^2 \mathrm{s}^{-1}$]} &
\colhead{($\mu$m)} &
\colhead{}&
\colhead{(K)} &
\colhead{[cm/s$^{2}$]} &
\colhead{} &
\colhead{} &
\colhead{[$\mathrm{cm}^2 \mathrm{s}^{-1}$]} &
\colhead{($\mu$m)} &
\colhead{}
}
\startdata
\sidehead{\underline{2MASS 0045+16:}}
~~BT-Settl AGSS09 &2000&4.50&+0.5& \nodata& \nodata& \nodata & {253710} & 2000&5.00&+0.5& \nodata& \nodata& \nodata & {493972} \\
~~BT-Settl CIFIST11 &1800&5.00& 0.0\tablenotemark{a} & \nodata& \nodata& \nodata & {184374} & 1800&5.00& 0.0\tablenotemark{a} & \nodata& \nodata& \nodata & {202512} \\
~~ Drift-Phoenix &1700&4.00& 0.0\tablenotemark{a} & \nodata& \nodata& \nodata & {217628} & 1800&4.50& 0.0\tablenotemark{a} & \nodata& \nodata& \nodata & {359250} \\
~~Saumon \& Marley &1600&4.48& 0.0\tablenotemark{a}&1&6& \nodata & {174446} & 1600&4.48& 0.0\tablenotemark{a}&1&6& \nodata & {231808} \\
~~Madhusudhan &1700&4.0&0.0& \nodata & \nodata &100 & {485212} & 1700&4.0&0.0& \nodata & \nodata &100 & {612891} \\
~~Tremblin & 2000&3.5\tablenotemark{a}& 0.0\tablenotemark{a}& \nodata & 6\tablenotemark{a} & \nodata & {174357} & 2000&3.5\tablenotemark{a}&0.0\tablenotemark{a}& \nodata & 6\tablenotemark{a} &\nodata & {366177} \\
\sidehead{\underline{G196$-$3B:}}
~~BT-Settl AGSS09 &1700&3.50&0.0& \nodata& \nodata& \nodata & {15589} & 1700&4.50&0.0& \nodata& \nodata& \nodata & {42102} \\
~~BT-Settl CIFIST11 &1750&4.00& 0.0\tablenotemark{a} & \nodata& \nodata& \nodata & {14745} & 1700&3.50& 0.0\tablenotemark{a} & \nodata& \nodata& \nodata & {40166} \\
~~ Drift-Phoenix &1600&4.50& 0.0\tablenotemark{a} & \nodata& \nodata& \nodata & {28231} & 1600&4.00& 0.0\tablenotemark{a} & \nodata& \nodata& \nodata & {43899} \\
~~Saumon \& Marley &1400&4.25& 0.0\tablenotemark{a}&1&4& \nodata & {21832} & 1400&4.25& 0.0\tablenotemark{a}&1&4& \nodata & {49006} \\
~~Madhusudhan &1400&4.0&0.0& \nodata & \nodata &60 & {47408} & 1400&4.0&0.0& \nodata & \nodata &100 & {103123} \\
~~Tremblin & 1600&3.5\tablenotemark{a}& 0.0\tablenotemark{a}& \nodata & 6\tablenotemark{a} & \nodata & {15486} & 1600&3.5\tablenotemark{a}&0.0\tablenotemark{a}& \nodata & 6\tablenotemark{a} &\nodata & {23985} \\
\sidehead{\underline{2MASS 0501$-$00:}}
~~BT-Settl AGSS09 &2000&5.50&+0.5& \nodata& \nodata& \nodata & {23214} & 1700&4.50&0.0& \nodata& \nodata& \nodata & {105845} \\
~~BT-Settl CIFIST11 &1650&5.00& 0.0\tablenotemark{a} & \nodata& \nodata& \nodata & {18834} & 1600&5.00& 0.0\tablenotemark{a} & \nodata& \nodata& \nodata & {106892} \\
~~ Drift-Phoenix &1600&3.50& 0.0\tablenotemark{a} & \nodata& \nodata& \nodata & {18862} & 1600&3.50& 0.0\tablenotemark{a} & \nodata& \nodata& \nodata & {35401} \\
~~Saumon \& Marley &1400&4.00& 0.0\tablenotemark{a}&1&6& \nodata & {15106} & 1400&4.25& 0.0\tablenotemark{a}&1&4& \nodata & {106145} \\
~~Madhusudhan &1500&4.0&0.0& \nodata & \nodata &60 & {53237} & 1400&4.0&0.0& \nodata & \nodata &100 & {222405} \\
~~Tremblin & 1800&3.5\tablenotemark{a}& 0.0\tablenotemark{a}& \nodata & 6\tablenotemark{a} & \nodata & {12864} & 1600&3.5\tablenotemark{a}&0.0\tablenotemark{a}& \nodata & 6\tablenotemark{a} &\nodata & {70538} \\
\sidehead{\underline{2MASS 0355+11:}}
~~BT-Settl AGSS09 &1700&4.50&0.0& \nodata& \nodata& \nodata & {13796} & 1600&3.50&0.0& \nodata& \nodata& \nodata & {75661} \\
~~BT-Settl CIFIST11 &1550&4.00& 0.0\tablenotemark{a} & \nodata& \nodata& \nodata & {5305} & 1550&4.00& 0.0\tablenotemark{a} & \nodata& \nodata& \nodata & {21312} \\
~~ Drift-Phoenix &1500&3.50& 0.0\tablenotemark{a} & \nodata& \nodata& \nodata & {11847} & 1600&5.00& 0.0\tablenotemark{a} & \nodata& \nodata& \nodata & {94024} \\
~~Saumon \& Marley &1200&4.25& 0.0\tablenotemark{a}&1&6& \nodata & {12298} & 1200&5.47& 0.0\tablenotemark{a}&1&4& \nodata & {77846} \\
~~Madhusudhan &1300&3.75&0.0& \nodata & \nodata &60 & {17793} & 1200&4.25&0.0& \nodata & \nodata &60 & {187057} \\
~~Tremblin & 1400&3.5\tablenotemark{a}& +0.2\tablenotemark{a}& \nodata & 6\tablenotemark{a} & \nodata & {7631} & 1400&3.5\tablenotemark{a}&0.0\tablenotemark{a}& \nodata & 6\tablenotemark{a} &\nodata & {67283} \\
\sidehead{\underline{2MASS 0103+19:}}
~~BT-Settl AGSS09 &1500&4.50&0.0& \nodata& \nodata& \nodata & {5615} & 1700&4.50&0.0& \nodata& \nodata& \nodata & {24475} \\
~~BT-Settl CIFIST11 &1650&5.00& 0.0\tablenotemark{a} & \nodata& \nodata& \nodata & {2912} & 1500&5.00& 0.0\tablenotemark{a} & \nodata& \nodata& \nodata & {15322} \\
~~ Drift-Phoenix &1600&3.50& 0.0\tablenotemark{a} & \nodata& \nodata& \nodata & {4581} & 1600&4.50& 0.0\tablenotemark{a} & \nodata& \nodata& \nodata & {11271} \\
~~Saumon \& Marley &1400&4.25& 0.0\tablenotemark{a}&1&6& \nodata & {2551} & 1300&4.00& 0.0\tablenotemark{a}&1&2& \nodata & {17097} \\
~~Madhusudhan &1400&4.0&0.0& \nodata & \nodata &60 & {7222} & 1300&4.0&0.0& \nodata & \nodata &100 & {28273} \\
~~Tremblin & 1800&3.5\tablenotemark{a}& 0.0\tablenotemark{a}& \nodata & 6\tablenotemark{a} & \nodata & {4101} & 1600&3.5\tablenotemark{a}&0.0\tablenotemark{a}& \nodata & 6\tablenotemark{a} &\nodata & {17127} \\
\sidehead{\underline{2MASS 2244+20:}}
~~BT-Settl AGSS09 &1700&4.50&0.0& \nodata& \nodata& \nodata & {47015} & 1600&3.50&0.0& \nodata& \nodata& \nodata & {109410} \\
~~BT-Settl CIFIST11 &1400&4.50& 0.0\tablenotemark{a} & \nodata& \nodata& \nodata & {25443} & 1400&4.00& 0.0\tablenotemark{a} & \nodata& \nodata& \nodata & {58998} \\
~~ Drift-Phoenix &1500&3.50& 0.0\tablenotemark{a} & \nodata& \nodata& \nodata & {47917} & 1500&3.50& 0.0\tablenotemark{a} & \nodata& \nodata& \nodata & {63373} \\
~~Saumon \& Marley &1200&4.00& 0.0\tablenotemark{a}&1&4& \nodata & {10359} & 1000&4.00& 0.0\tablenotemark{a}&1&6& \nodata & {27231} \\
~~Madhusudhan &1300&4.0&0.0& \nodata & \nodata &60 & {44486} & 1200&4.0&0.0& \nodata & \nodata &60 & {78465} \\
~~Tremblin & 1400&3.5\tablenotemark{a}& +0.2\tablenotemark{a}& \nodata & 6\tablenotemark{a} & \nodata & {15470} & 1400&3.5\tablenotemark{a}&+0.2\tablenotemark{a}& \nodata & 6\tablenotemark{a} &\nodata & {40499} \\
\sidehead{\underline{PSO 318.5$-$22:}}
~~BT-Settl AGSS09 &1600&3.50&0.0& \nodata& \nodata& \nodata & {3633} & 1600&3.50&0.0& \nodata& \nodata& \nodata & {54994} \\
~~BT-Settl CIFIST11 &1350&3.50& 0.0\tablenotemark{a} & \nodata& \nodata& \nodata & {2861} & 1550&3.50& 0.0\tablenotemark{a} & \nodata& \nodata& \nodata & {11876} \\
~~ Drift-Phoenix &1500&5.00& 0.0\tablenotemark{a} & \nodata& \nodata& \nodata & {5423} & 1500&5.00& 0.0\tablenotemark{a} & \nodata& \nodata& \nodata & {9482} \\
~~Saumon \& Marley &1000&4.25& 0.0\tablenotemark{a}&1&6& \nodata & {2122} & 900&5.00& 0.0\tablenotemark{a}&1&4& \nodata & {8704} \\
~~Madhusudhan &1100&4.0&0.0& \nodata & \nodata &30 & {5440} & 1000&4.0&+0.5& \nodata & \nodata &60 & {21581} \\
~~Tremblin & 1275&3.7\tablenotemark{a}& +0.4\tablenotemark{a}& \nodata & 5\tablenotemark{a} & \nodata & {1694} & 1200&3.5\tablenotemark{a}&0.0\tablenotemark{a}& \nodata & 6\tablenotemark{a} &\nodata & {12564} \\
\sidehead{\underline{WISE 0047+68:}}
~~BT-Settl AGSS09 &1700&4.50&0.0& \nodata& \nodata& \nodata & {23472} & 1600&3.50&0.0& \nodata& \nodata& \nodata & {131724} \\
~~BT-Settl CIFIST11 &1400&4.00& 0.0\tablenotemark{a} & \nodata& \nodata& \nodata & {13212} & 1500&3.50& 0.0\tablenotemark{a} & \nodata& \nodata& \nodata & {57278} \\
~~ Drift-Phoenix &1500&3.50& 0.0\tablenotemark{a}& \nodata& \nodata& \nodata & {25152} & 1500&4.00& 0.0\tablenotemark{a}& \nodata& \nodata& \nodata & {47773} \\
~~Saumon \& Marley &1100&4.47& 0.0\tablenotemark{a}&1&6& \nodata & {5711} & 1000&4.00& 0.0\tablenotemark{a}&1&6& \nodata & {20615} \\
~~Madhusudhan &1200&4.25&0.0& \nodata & \nodata &60 & {19659} & 1100&4.0&+0.5& \nodata & \nodata &100 & {76802}\\
~~Tremblin & 1400&3.5\tablenotemark{a}& +0.2\tablenotemark{a}& \nodata & 6\tablenotemark{a} & \nodata & {5935} & 1275&3.7\tablenotemark{a}&+0.4\tablenotemark{a}& \nodata & 5\tablenotemark{a} &\nodata & {25983} \\
\enddata
\tablenotetext{a}{Not a free parameter in the fit}
\end{deluxetable}
\end{landscape}

\begin{figure*}
\includegraphics[trim = {.25cm 4.7cm 3.5cm 2.75cm},clip,width=.9\textwidth]{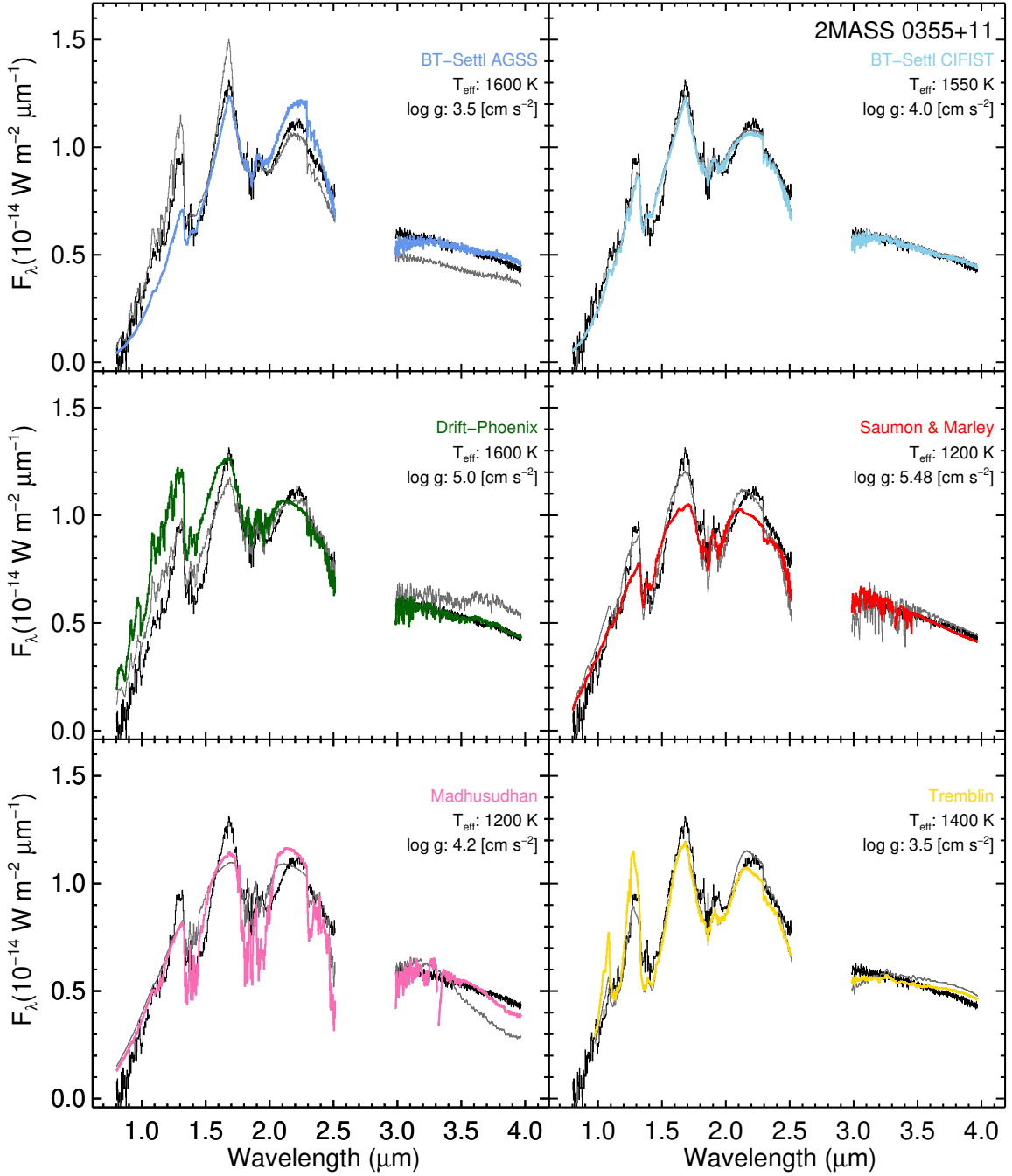}
\centering
\caption{The combined spectrum of 2MASS 0355+11 (black) compared to the model spectra (colored) with parameters that best fits the entire combined spectrum. For ease of comparison, the near-IR best-fit models are shown in grey.}\label{fig:0355Fits}
\end{figure*}

\begin{figure*}
\includegraphics[trim = {.25cm 4.7cm 3.5cm 2.75cm},clip,width=.9\textwidth]{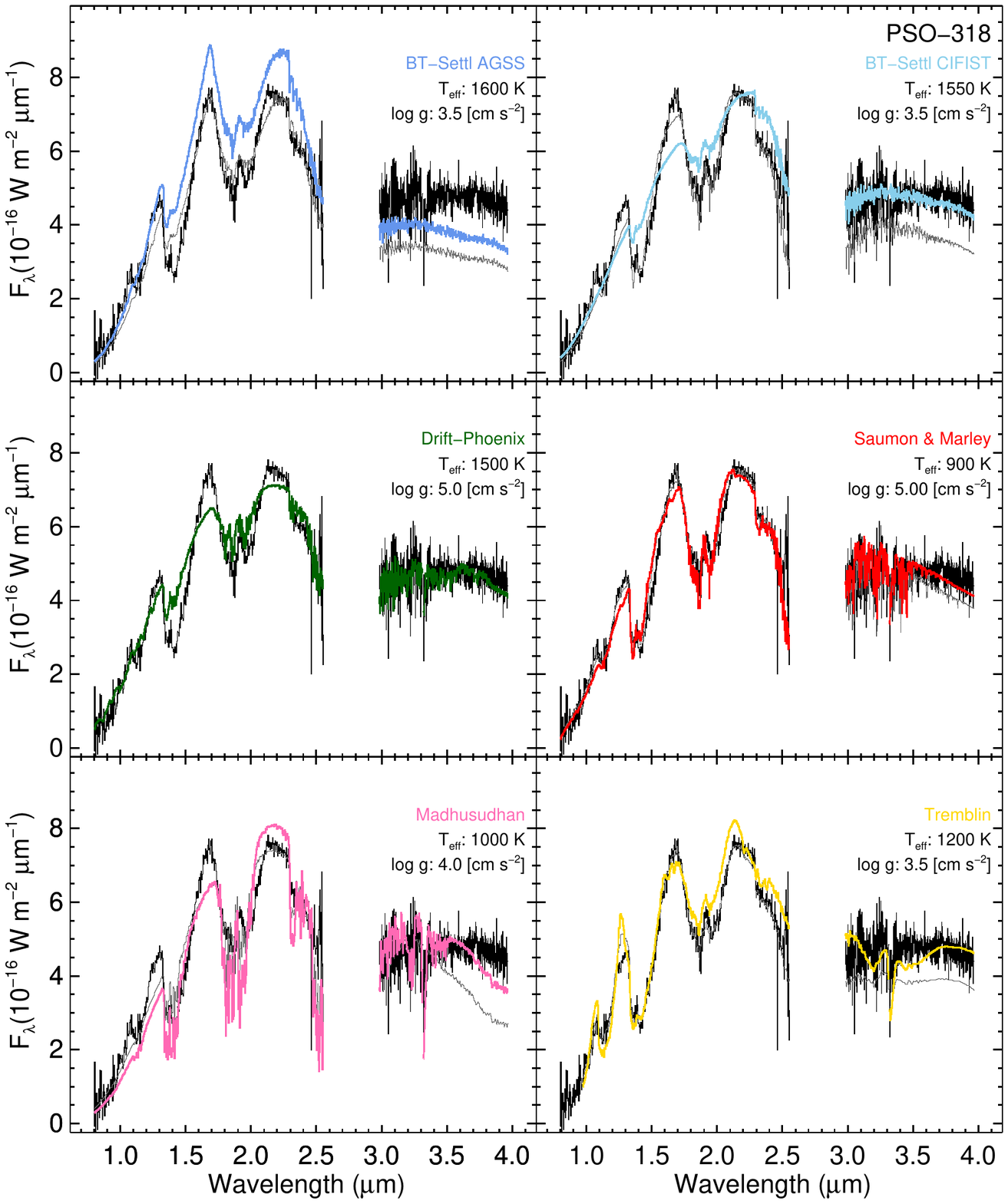}
\centering
\caption{Same as Figure \ref{fig:0355Fits}, except using the cooler PSO 318.5$-$22}
\label{fig:318Fits}
\end{figure*}

\begin{figure*}
\includegraphics[trim = {0 5cm 1cm 8.5cm}, clip,width=.75\textwidth]{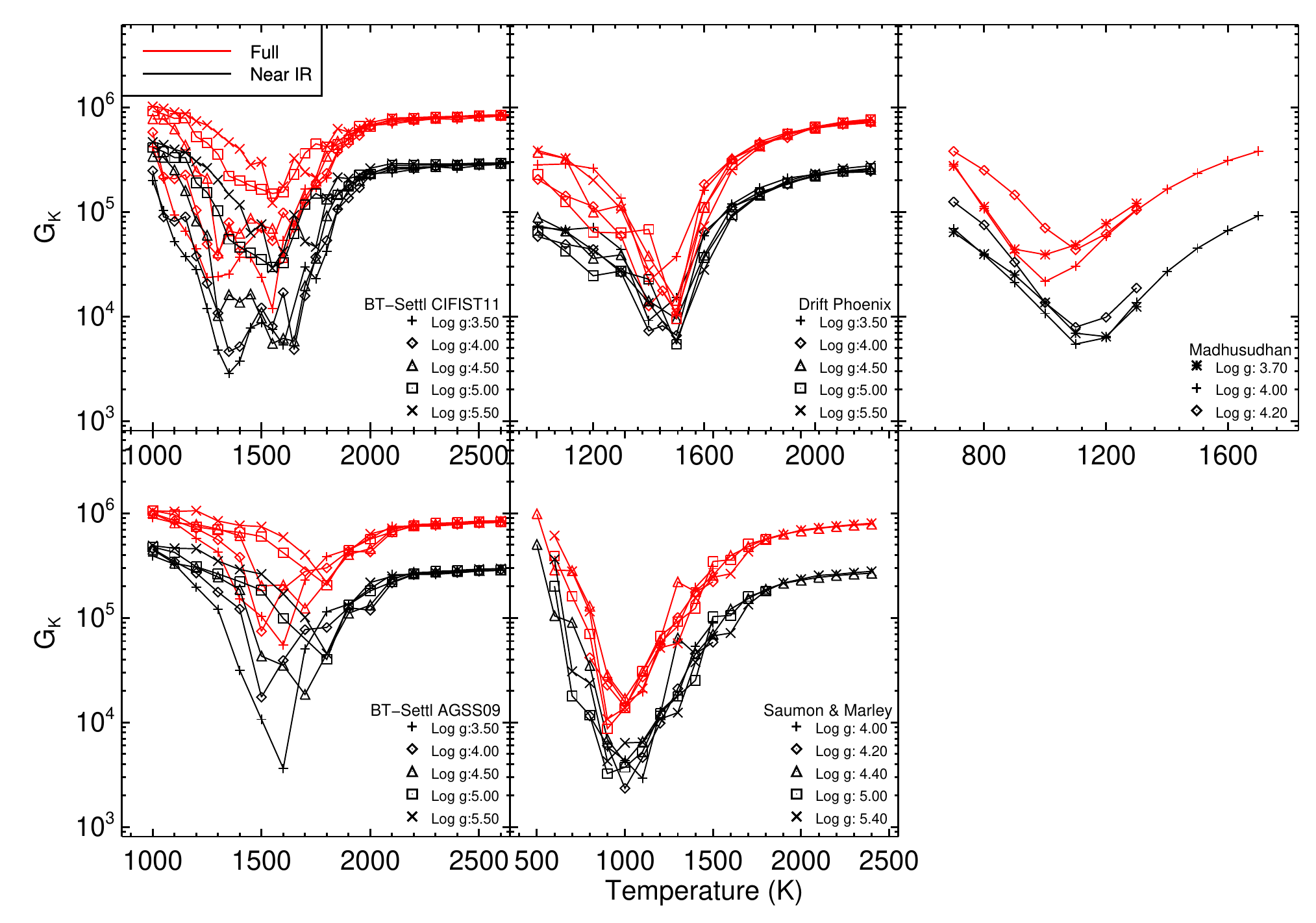}
\centering
\caption{The goodness-of-fit parameter (\gk) for PSO 318.5$-$22 plotted as function of temperature and \logg{}. Note that the shape does not change drastically when including the $L$ band, but it does move to cooler temperatures ($\sim$100 K). This held true across all of our objects.}
\label{fig:GK}
\end{figure*}

\subsubsection{Mixing Rate}

We can see how the vertical mixing rate (\kzz) tends to vary with the inclusion of the $L$ band through the Saumon \& Marley fits. With the inclusion of the $L$-band spectra, there is a shift to lower \kzz~values (from the max possible 6 [$\mathrm{cm}^2 \mathrm{s}^{-1}$] to 4 or even 2) for the combined spectra, which on face value disagrees with the conclusions from \citet{2018ApJ...869...18M} finding a need for a high \kzz~(around $10^8$ $\mathrm{cm}^2 \mathrm{s}^{-1}$) to explain $L$-band features. However, when the models' thickest cloud parameter was used (f1), as was the case in the best fit for all of our objects, changing the log \kzz~value from 6 to 4 had little affect on the model spectra in the near-IR and $L$ band. The effect on \gk~was at its highest a 0.5 percent increase, and sometimes was as small as 0.01 percent. For some temperatures, the insensitivity of the model to \kzz~extends down to \kzz~of 2 [$\mathrm{cm}^2 \mathrm{s}^{-1}$], like at the best-fit temperature of 2MASS 0103+19 (1300 K), but often the lowered vertical mixing would start to deepen the $L$-band methane features at this point, making the fit noticeably worse. When a thinner cloud parameter was set, the best-fit \kzz~values tended to stay high. Overall, we have a low sensitivity between the higher \kzz~values, but can still say we agree with \citet{2018ApJ...869...18M} that higher values in general create better fits. 

\subsection{Model Comparisons: Evolutionary Parameters} \label{EvoParam}
\begin{figure*}
\includegraphics[trim = {.25cm 4.7cm 3.5cm 2.75cm},clip,width=.9\textwidth]{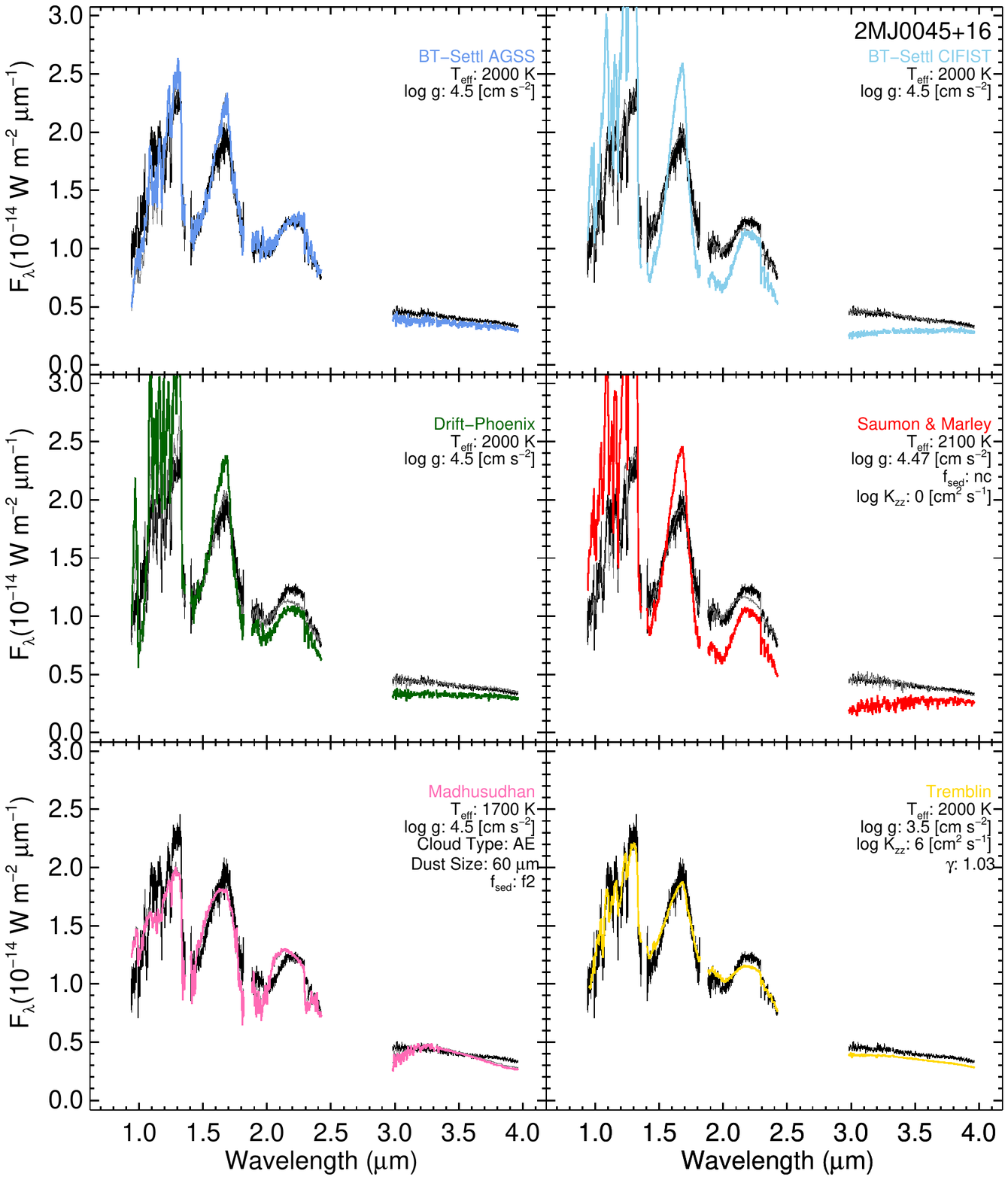}
\centering
\caption{The spectra of 2MASS 0045+16 compared to the spectra from each model set that best matches the known evolutionary effective temperature and surface gravity found in Table \ref{tbl:EVO}. Note that in this case the Tremblin models do not extend to 4.5 [$\mathrm{cm}^2 \mathrm{s}^{-1}$] and the Madhusudhan models do not extend to 2000 K. In gray is also the combined best-fit spectra (in the case of the Tremblin models, the best-fit and evolutionary model parameters are the same).}
\label{fig:0045_EVO}
\end{figure*}

\begin{figure*}
\includegraphics[trim = {.25cm 4.7cm 3.5cm 2.75cm},clip,width=.9\textwidth]{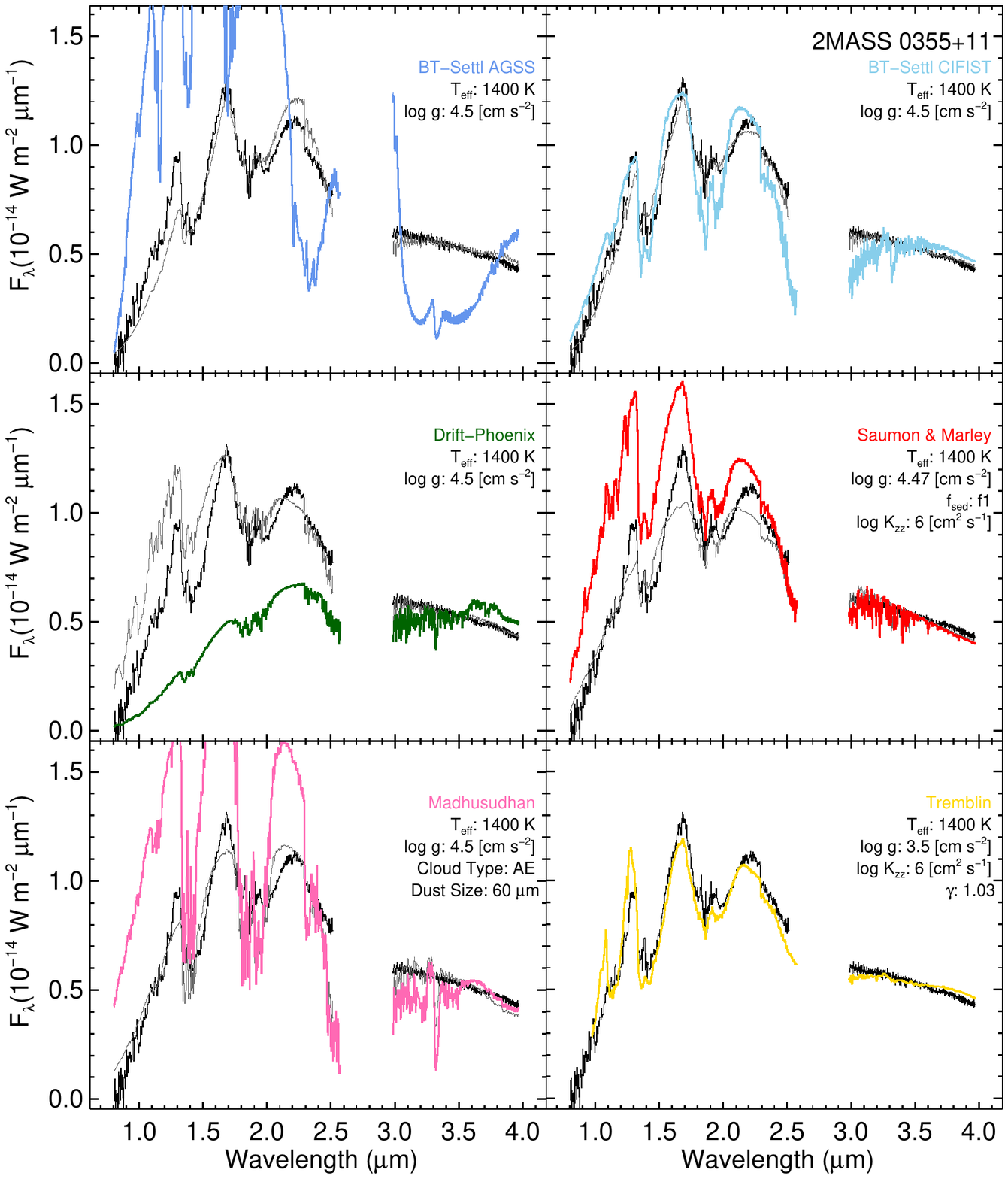}
\centering
\caption{The spectra of 2MASS 0355+11 compared to the spectra from each model set that best matches the known evolutionary effective temperature and surface gravity found in Table \ref{tbl:EVO}. Note that in this case the Tremblin models do not extend to 4.5 [$\mathrm{cm}^2 \mathrm{s}^{-1}$]. In gray is also the combined best-fit spectra (in the case of the Tremblin models, the best-fit and evolutionary model parameters are the same).}
\label{fig:0355_EVO}
\end{figure*}

\begin{figure*}
\includegraphics[trim = {.25cm 4.7cm 3.5cm 2.75cm},clip,width=.9\textwidth]{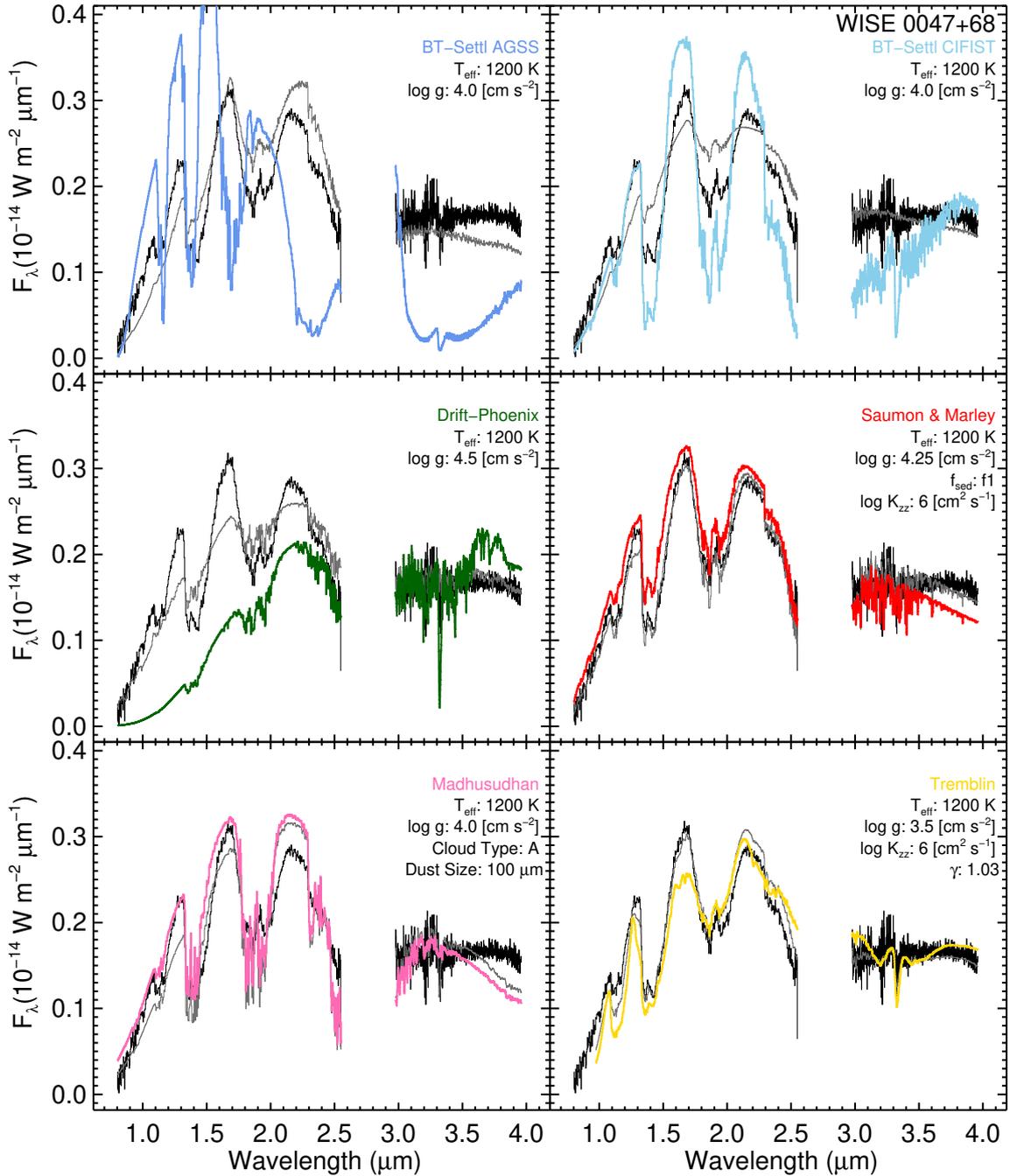}
\centering
\caption{The spectra of WISE 0047+68 compared to the spectra from each model set that best matches the known evolutionary effective temperature and surface gravity found in Table \ref{tbl:EVO}. Note that in this case the Tremblin models do not extend to 4.0 [$\mathrm{cm}^2 \mathrm{s}^{-1}$]. In gray is also the combined best-fit spectra.}
\label{fig:0047_EVO}
\end{figure*}

\begin{figure*}
\includegraphics[trim = {.25cm 4.7cm 3.5cm 2.75cm},clip,width=.9\textwidth]{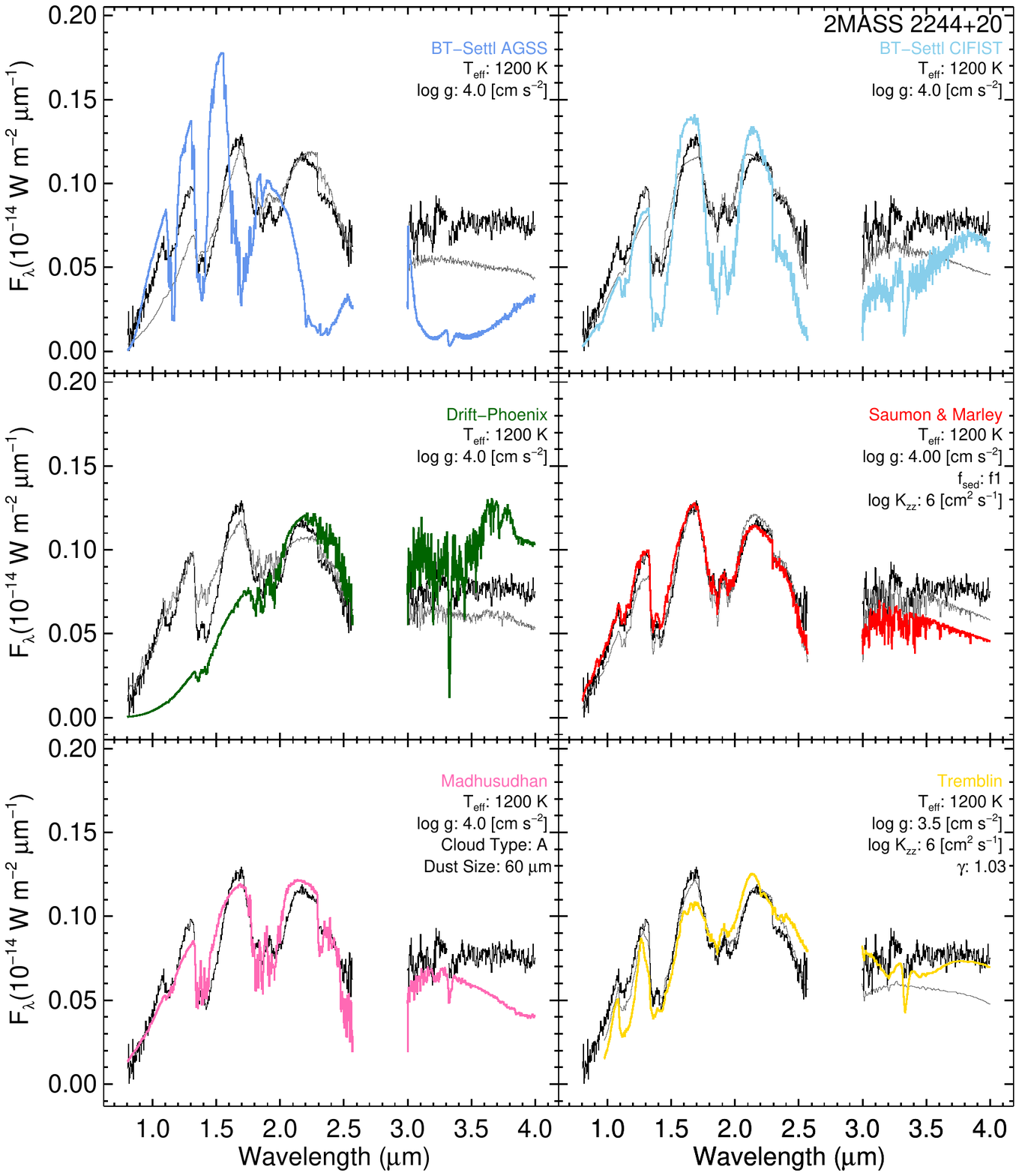}
\centering
\caption{The spectra of 2MASS 2244+20 compared to the spectra from each model set that best matches the known evolutionary effective temperature and surface gravity found in Table \ref{tbl:EVO}. Note that in this case the Tremblin models do not extend to 4.0 [$\mathrm{cm}^2 \mathrm{s}^{-1}$]. In gray is also the combined best-fit spectra (in the case of the Madhusudhan models, the best-fit and evolutionary model parameters are the same).}
\label{fig:2244_EVO}
\end{figure*}

\begin{figure*}
\includegraphics[trim = {.25cm 4.7cm 3.5cm 2.75cm},clip,width=.9\textwidth]{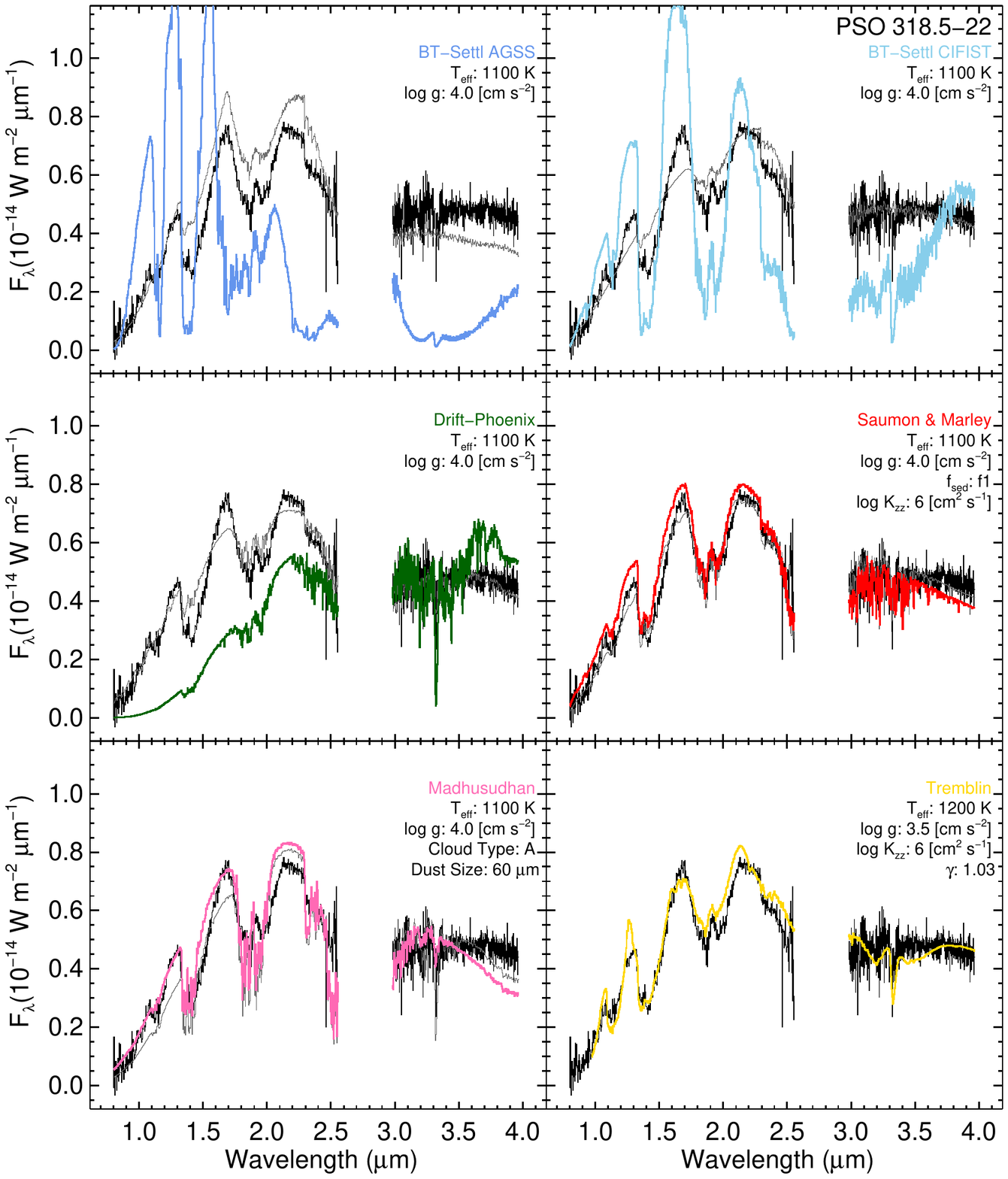}
\centering
\caption{The spectra of PSO 318.5$-$22 compared to the spectra from each model set that best matches the known evolutionary effective temperature and surface gravity found in Table \ref{tbl:EVO}. Note that in this case the Tremblin models do not extend to 1100 K or 4.0 [$\mathrm{cm}^2 \mathrm{s}^{-1}$]. In gray is also the combined best-fit spectra (in the case of the Tremblin models, the best-fit and evolutionary model parameters are the same).}
\label{fig:PSO_EVO}
\end{figure*}

Some of our objects are members of young moving groups and therefore have ages associated with them. These ages, along with measured $L_{\mathrm{bol}}$ and evolutionary models, give well-defined evolutionary parameters for five of our objects, 2MASS 0045+16, 2MASS 2244+20, WISE 0047+68, 2MASS 0355+11 \citep[All from ][]{2016ApJS..225...10F} and PSO 318.5$-$22 \citep{2016ApJ...819..133A}. The memberships, ages, masses, effective temperature, and surface gravities for all of these objects can be found in Table \ref{tbl:EVO}. We can us these \teff~and \logg{} values to select synthetic spectra that match these parameters, which we'll hereafter refer to as evolutionary effective temperature and surface gravity. For each model set, we took the model spectrum with the effective temperature and surface gravity closest to each object's evolutionary effective temperature and surface gravity, and of these selected the model parameters and $C$ which minimized \gk. The Madhusudhan and Tremblin models do not fully cover the range of temperatures and gravities needed to encompass all of the objects evolutionary temperatures and gravities. When this occurred, the closest available parameters were used. The comparisons for these objects can be seen in Figures \ref{fig:0045_EVO} through \ref{fig:PSO_EVO}. For ease of reference, the combined best-fit model from the previous section is also plotted in grey.

\begin{table*}
\caption{Our Sample of Young L Dwarfs With Known Moving Groups and Ages} \label{tbl:EVO}
\begin{tabular}{lcccccc}
\toprule[1.5pt]
Name & YMG & Age & Mass Range & \teff & \logg{} & Refs \\
 & Membership & (Myrs) & ($M_\mathrm{Jup}$) & (K) & [$\mathrm{cm}^2 \mathrm{s}^{-1}$] & \\
\toprule[1.5pt]
2MASS J0045+16	&	Argus   &  40 & 20--29& $2059 \pm 45$& $4.22 \pm 0.10$ &	C09, F15, L16, F16	\\
2MASS J0355+11	&	AB Dor	& 110--130  & 15--27& $1478\pm 58$  & $4.58^{+0.07}_{-0.17}$ &	C09, F13, AL13, F16	\\
WISEP J0047+68 &	AB~Dor  & 110--130  & 9--15	& $1230 \pm 27$& $4.21 \pm 0.10$ &	G15, F15 ,F16\\
2MASS J2244+20	&	AB~Dor	& 110--130  & 9--12 &$1184 \pm 10$& $4.18 \pm 0.08$ &	K08, F15, A16 	\\
PSO J318.5$-$22	&$\beta$~Pic& 20--26  & 5--7  & $1127^{+24}_{-26}$ & 4.01 $\pm$ 0.03 &	L13, A16, F16	
\end{tabular}
\tablerefs{(A16)~\citet{2016ApJ...819..133A}, (AL13)~\citet{2013ApJ...772...79A},
(C09)~\citet{2009AJ....137.3345C}, 
(F12)~\citet{2012ApJ...752...56F}, (F13)~\citet{2013AJ....145....2F},
(F15)~\citet{2015ApJ...810..158F}
(F16)~\citet{2016ApJS..225...10F},
(G15)~\citet{2015ApJ...799..203G},
(K08)~\citet{2008ApJ...689.1295K},
(L16)~\citet{2016ApJ...833...96L}, (L13)~\citet{2013ApJ...777L..20L},
(V18)~\citet{2018MNRAS.474.1041V}
}

\end{table*}
For all our objects except 2MASS 0045+16, we see that the Drift-Phoenix and the BT-Settl AGSS models do not fit well when the evolutionary effective temperature and surface gravity are used. The Drift-Phoenix model over-predicts the flux in the $L$ band, with the peaks of the $J$ and $H$ bands being almost non-existent, which can be most clearly seen in Figure \ref{fig:2244_EVO}. The AGSS model has the opposite problem, as already at 1400 K in Figure \ref{fig:0355_EVO} it looks remarkably like a T dwarf with its deep methane absorption in the near-IR and $L$ band. The fact that the methane absorption is this strong in the AGSS model at such high temperatures is part of why the best-fit temperatures were hot for our young L dwarfs. The exception is our hottest object 2MASS 0045+16 (Figure \ref{fig:0045_EVO}), which the AGSS model arguably has the best fit. Meanwhile, the Drift-Phoenix model inverts its color with the $J$ and $H$ band being too strong compared to weaker $L$ and $K$ bands, though not quite as dramatically at this temperature.

The CIFIST model set is one of the better fits for 2MASS 0355+11 when using the evolutionary effective temperature and surface gravity, where its biggest issues are predicting stronger water absorption features than we observe, both between the near-IR bands and in the $L$ band, and missing the sharp triangle shape of the H band. This issue is there in all the objects, but it is most pronounced for the colder ones like in Figure \ref{fig:0047_EVO}, where the predicted water absorption becomes even stronger, and is accompanied by strong methane features as well. This increased absorption drops the model's $L$-band flux level too low. In general, the CIFIST BT-Settl model set seems to do comparatively well for young L dwarfs at higher temperatures, but struggles when transitioning to lower temperatures. 

The selected Madhusudhan models have the opposite problem, fitting colder objects better than hotter ones. When using the evolutionary effective temperature and surface gravity of 2MASS 0355+11 and 2MASS 0045+16, we see the differences in the flux levels of the near-IR and $L$ band are much greater in the model than in the observed spectrum. This disparity is not there for the three colder objects, though other issues occur, such as over-estimation of the $K$-band flux and a turndown at the long end of the $L$ band. However, these issues are minor compared to the flux level spread between the bands at higher temperatures, which explains the Madhusudhan model’s tendency to give colder best-fit temperatures.

The Tremblin and Saumon \& Marley model sets are the most consistently good fits, though both still have some discrepancies. The Saumon \& Marley models tend to overestimate the flux in the $J$ and $H$ bands, especially for the 2MASS 0045+16 and 2MASS 0035+11. Models with lower temperatures have lower near-IR flux which are closer to the observed values, such as in Figure \ref{fig:0047_EVO}, which explains why the models' best fits trend towards colder temperatures. The Tremblin models are off in the near-IR about as much as the Saumon \& Marley models, but with no consistent issues. For the three coldest objects, the Tremblin models matches the $L$-band flux level, yet exhibits strong absorption in the $P$, $Q$, and $R$ branches of methane unseen in the observed spectrum. Both also under-predict the $L$-band flux for 2MASS 0045+16, though this is consistent across all the models.

Looking more generally, the three of the four models without disequilibrium chemistry (both BT-Settl models and  Drift-Phoenix) show an onset of methane at higher temperatures than we observe, such as in Figure \ref{fig:0355_EVO}. This agrees with what we saw from the best-fit models. This effect is especially notable in the BT-Settl AGSS models, whose spectra at this temperature look more similar to T dwarfs than L dwarfs. The Madhusudhan model with the thickest clouds (Type A) does not have this early appearance of methane, but the less thick AE clouds do, with $Q$-band absorption being strong even at 1400 K. As for the models with disequilibrium chemistry, the $Q$-branch of methane is well fit when using the evolutionary effective temperature and surface gravity, and overall are generally the better fit compared to the equilibrium models. We also find that the cloudless BT-Settl models clouds tend to fit the colder objects worse than those with clouds as their evolutionary temperature, as seen in Figures \ref{fig:0047_EVO}-\ref{fig:PSO_EVO}. The exception to the cloudy models fitting better is the Drift-Phoenix models, which are poor past 1500 K due to so little flux coming out in the $J$ and $K$ bands. This is evidence that some combination of clouds and disequilibrium chemistry is a necessary for modeling brown dwarf atmospheres with accurate effective temperatures.

\subsection{AB Dor Sub-population}
\begin{figure*}
\includegraphics[width = \textwidth]{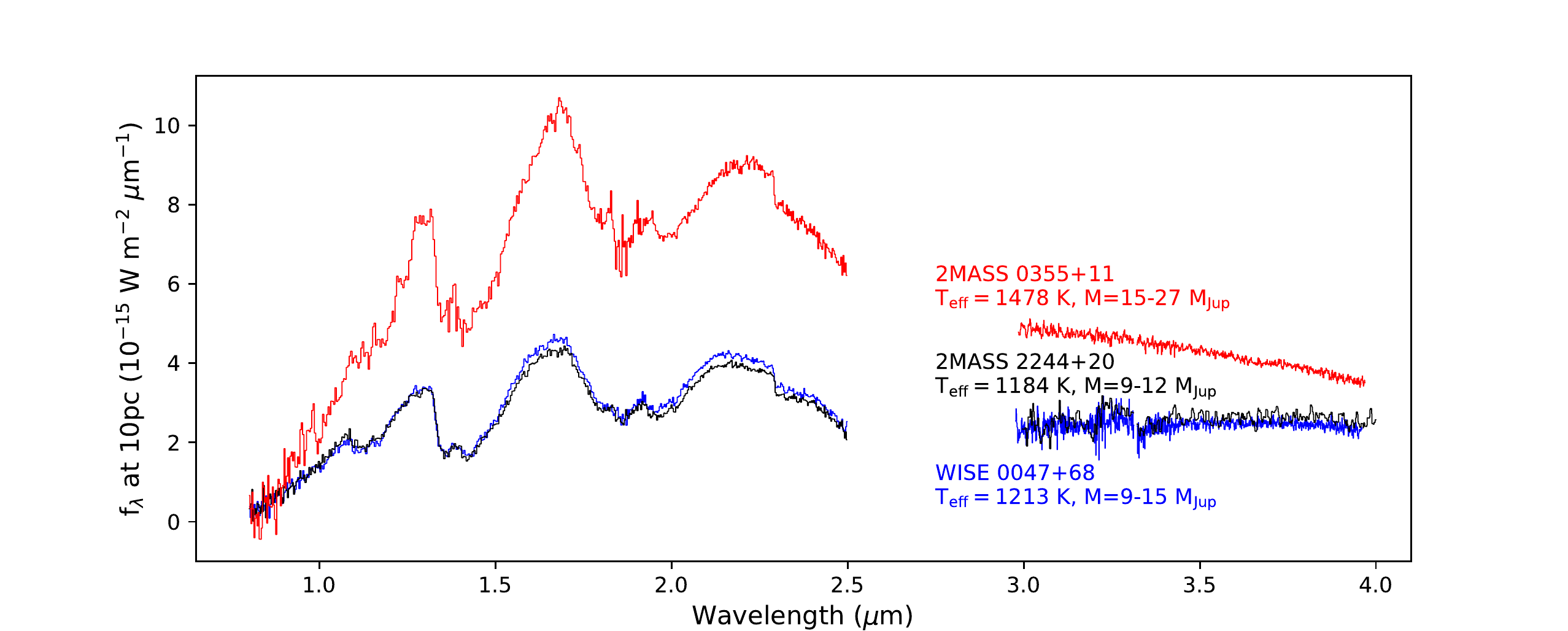}
\centering
\caption{Spectra of three members of the young moving group AB Dor, placed at 10 pc. The two objects of the same mass lie almost on top of each other, and with a $Q$-branch methane feature of similar depth. This feature is absent in the hotter and more massive 2MASS 0355+11.}\label{fig:ABDOR}
\end{figure*}

Three of our targets, 2MASS 2244+20, WISE 0047+68, and 2MASS 0355+13, are members of the AB Doradus young moving group, and thus can be assumed to have similar compositions and ages \citep[$\sim$120 Myrs,][]{2013ApJ...766....6B} due to their shared origin and formation history. WISE 0047+68 and 2MASS 2244+20 have approximately the same mass, respectively 9-15 and 9-12 \mjup{}, but 2MASS 0355+13 is almost twice as massive at 15-27 \mjup. This means Figure \ref{fig:ABDOR}, which shows the spectra of these three objects, is a mass sequence at a fixed age and composition, though admittedly an abbreviated one. 2MASS 0355+13 is more luminous than WISE 0047+68 and 2MASS 2244+20, whose spectra lie nearly on top of each other due to having similar masses (See Table \ref{tbl:EVO}). Of particular interest is the 3.3 $\mu$m $Q$-branch methane feature, which has a similar depth in the two cooler objects, even with different infrared spectral types and gravity classification (WISE 0047+68 as L7 INT-G and 2MASS 2244+20 as L6 VL-G). This feature is absent in the more massive (and hotter) 2MASS 0355+13, showing the temperature dependence of the methane feature for a fixed age and composition. AB Dor has several other ultracool members \citep{2016ApJS..225...10F}, which means it will be possible to expand this mass sequence.  

\section{Conclusions}\label{sec:con}

From this work, we can see that $L$-band spectra are not particularly strong diagnostic tools on their own for young L dwarfs due to a lack of strong features (with the exception of the $Q$-branch of methane). However, when used in hand with near-IR spectra it is extremely useful for understanding objects' atmospheric properties. The inclusion of the $L$ band highlights where models are falling short, and what physical processes models need to include to match observations. We find adding the $L$ band lowers the best-fit effective temperature of our models by $\sim$100 K. For these fits, when clouds are included in the models, thick ones are preferred, and when a vertical mixing rate is used (\kzz), higher values give the best results. We also note that for models to fit the combined spectra well at the published evolutionary effective temperatures for these objects, we need models with disequilibrium chemistry and/or clouds. Overall, we find the Tremblin and Saumon \& Marley models fit the full spectrum of the young objects best. These results show the power of wide spectral coverage, matching conclusion from \citet{2021MNRAS.506.1944B}, and we recommend that our data also be used to enhance future retrievals of these objects to parse their diversity. In particular though, these observations show the value of the $L$ band for understanding the atmospheres of young brown dwarfs, and the giant exoplanets for which they act as proxies and give us a preview of the insights that the James Webb Space Telescope will bring. 

\section*{Acknowledgements}
Based on observations obtained at the Gemini Observatory, which is operated by the Association of Universities for Research in Astronomy, Inc., under a cooperative agreement with the NSF on behalf of the Gemini partnership: the National Science Foundation (United States), the National Research Council (Canada), CONICYT (Chile), Ministerio de Ciencia, Tecnología e Innovación Productiva (Argentina), and Ministério da Ciência, Tecnologia e Inovação (Brazil).  KNA and SAB are grateful for support from the Isaac J. Tressler fund for astronomy at Bucknell University, and the Doreen and Lyman Spitzer Graduate Fellowship in Astrophysics at the University of Toledo. This work benefited from the Exoplanet Summer Program in the Other Worlds Laboratory (OWL) at the University of California, Santa Cruz, a program funded by the Heising-Simons Foundation. We appreciated conversations with Brittney Miles which enhanced this work, and we would like to thank Denise Stephens for contributing the 2MASS 2244+20 spectrum, and Pascal Tremblin for allowing us access to unpublished models. This work also benefited from The UltracoolSheet, maintained by Will Best, Trent Dupuy, Michael Liu, Rob Siverd, and Zhoujian Zhang, and developed from compilations by Dupuy \& Liu (2012, ApJS, 201, 19), Dupuy \& Kraus (2013, Science, 341, 1492), Liu et al. (2016, ApJ, 833, 96), Best et al. (2018, ApJS, 234, 1), and Best et al. (2020b, AJ, in press).

\section*{Data Availability}
The combined spectra for all of these objects are available in the online supplementary material for this article.



\bibliographystyle{mnras}
\bibliography{example} 




\appendix
\section{IRAC Channel 1 Photometry}\label{app:irac}
PSO 318.5$-$22 does not have [3.6] magnitudes, so we used its spectral type, W1 magnitude \citep{2013ApJ...772...79A,2013ApJ...777L..20L} and a spectral type vs.~W1$-$[3.6] color relation to compute its [3.6] magnitude. The spectral type vs.~W1$-$[3.6] color relation was created from a linear least-squares fit to the values for young L dwarfs listed in Table \ref{tbl:irac}. We found the relation to be W1$-$[3.6] = $0.0822*$SpT$+0.015$, where SpT = 10 for a dwarf with a near-IR spectral type of L0, with a covariance matrix of:
$\begin{bmatrix}
1.18 \times 10^{-3} & -8.86 \times 10^{-5} \\
-8.86 \times 10^{-5} & 6.94 \times 10^{-6} 
\end{bmatrix}$

Several of these objects are known low-gravity brown dwarfs which lack published IRAC photometry, but have available IRAC data in the Spitzer Heritage Archive.  We computed IRAC Channel 1 photometry from Post-BCD pipeline mosaicked images using the same method as \citet{2013Sci...341.1492D}.  We calculated the fluxes in a 3\farcs6 radius with a background annulus of 3\farcs6-8\farcs4.  Photometry was performed using the {\tt phot} routine from IDL Astronomy User's Library \citep{landsman}.  We applied aperture corrections as recommended by the IRAC data handbook.  We determined uncertainties on our aperture photometry using a Monte Carlo approach to account for uncertainties from associated post-BCD uncertainty images.  Our final photometric uncertainties also include published uncertainties for aperture corrections and flux zero points.  For comparison, we also calculated photometry for the young brown dwarfs in \citet{2009ApJ...703..399L} and found that our results agree with their published IRAC magnitudes to within our typical error of 0.02-0.03 mags.  Table \ref{tbl:irac} includes IRAC photometry for low-gravity L dwarfs, including objects in our sample.

\begin{table*}
\tablecaption{IRAC and WISE photometry of young L dwarfs} \label{tbl:irac}
\begin{tabular}{llrrl}
\toprule[1.5pt]
Name &
SpT &
AllWISE &
IRAC &
Refs. \\
 &
Near-IR &
W1 (mag) &
ch1 (mag) &
\\
\toprule[1.5pt]
2MASS J00325584$-$4405058	& L0 \textsc{vl-g}	& $12.839\pm0.024$	& $12.585\pm0.017$ &	This work, \citet{2013ApJ...772...79A} \\
WISEP J004701.06+680352.1	& L7 \textsc{int-g}	& $11.881\pm0.023$	& $11.518\pm0.017$ &	This work, \citet{2015ApJ...799..203G} \\
2MASS J010332.03+1935361	& L6 \textsc{int-g}	& $13.177\pm0.024$	& $12.930\pm0.020$ &	\citet{2015ApJ...799..154M}, \citet{2013ApJ...772...79A} \\
2MASS J01415823$-$4633574	& L0 \textsc{vl-g}	& $12.584\pm0.023$	& $12.360\pm0.020$ &	\citet{2009ApJ...703..399L}, \citet{2013ApJ...772...79A} \\
2MASS J02411151$-$0326587	& L1 \textsc{vl-g}	& $13.649\pm0.025$	& $13.390\pm0.020$ &	\citet{2009ApJ...703..399L}, \citet{2013ApJ...772...79A} \\
2MASS J03231002$-$4631237	& L0 \textsc{vl-g}	& $13.091\pm0.024$	& $12.840\pm0.020$ &	\citet{2009ApJ...703..399L}, \citet{2018AJ....155...34C} \\
2MASS J03552337+1133437	& L3 \textsc{vl-g}	& $10.526\pm0.024$	& $10.280\pm0.017$ &	This work, \citet{2013ApJ...772...79A} \\
2MASS J03572695$-$4417305	& L0 \textsc{int-g}	& $12.486\pm0.023$	& $12.210\pm0.020$ &	\citet{2009ApJ...703..399L}, \citet{2018AJ....155...34C} \\
2MASS J04433761+0002051	& L0 \textsc{vl-g}	& $10.831\pm0.023$	& $10.550\pm0.020$ &	\citet{2009ApJ...703..399L}, \citet{2013ApJ...772...79A} \\
2MASS J05012406$-$0010452	& L3 \textsc{vl-g}	& $12.050\pm0.023$	& $11.770\pm0.020$ &	\citet{2009ApJ...703..399L}, \citet{2013ApJ...772...79A} \\
2MASSI J0518461$-$275645	& L1 \textsc{vl-g}	& $13.043\pm0.023$	& $12.797\pm0.017$ &	This work, \citet{2013ApJ...772...79A} \\
2MASS J06085283$-$2753583	& L0 \textsc{vl-g}	& $11.975\pm0.024$	& $11.750\pm0.020$ &	\citet{2009ApJ...703..399L}, \citet{2013ApJ...772...79A} \\
G 196-3B	& L-3 \textsc{vl-g}	& $11.700\pm0.023$	& $11.660\pm0.020$ & \citet{2010ApJ...715.1408Z}, \citet{2013ApJ...772...79A} \\
2MASS J11193254$-$1137466AB	& L7 \textsc{vl-g}	& $13.548\pm0.026$	& $13.234\pm0.017$ &	This work, \citet{2016ApJ...821L..15K} \\
WISEA J114724.10$-$204021.3	& L7 \textsc{vl-g}	& $13.718\pm0.026$	& $13.352\pm0.017$ &	This work, \citet{2016ApJ...822L...1S} \\
2MASSI J1615425+495321	& L3 \textsc{vl-g}	& $13.212\pm0.024$	& $12.910\pm0.020$ &	\citet{2009ApJ...703..399L}, \citet{2013ApJ...772...79A} \\
2MASS J17260007+1538190	& L3 \textsc{int-g}	& $13.067\pm0.024$	& $12.760\pm0.020$ &	\citet{2009ApJ...703..399L}, \citet{2013ApJ...772...79A} \\
PSO J318.5338$-$22.8603	& L7 \textsc{vl-g}	& $13.237\pm0.024$	& \nodata &	\citet{2013ApJ...777L..20L}\\
2MASS J2206449$-$421720	& L4 \textsc{vl-g}	& $12.856\pm0.023$	& $12.560\pm0.017$ &	This work, \citet{2018AJ....155...34C} \\
2MASS J22081363+2921215	& L3 \textsc{vl-g}	& $13.381\pm0.025$	& $13.080\pm0.020$ &	L09, \citet{2013ApJ...772...79A} \\
2MASS J22134491$-$2136079	& L0 \textsc{vl-g}	& $13.247\pm0.026$	& $12.990\pm0.020$ &	L09, \citet{2013ApJ...772...79A} \\
2MASS J22443167+2043433	& L6 \textsc{vl-g}	& $12.790\pm0.022$	& $12.350\pm0.020$ &	\citet{2007ApJ...655.1079L}, \citet{2013ApJ...772...79A} \\
SDSS J224953.47+004404.6AB	& L3 \textsc{int-g}	& $13.606\pm0.026$	& $13.340\pm0.020$ &	\citet{2015ApJ...799..154M}, \citet{2013ApJ...772...79A} \\
\end{tabular}
\end{table*}

\section{Combined Fits For The Remaining Objects}\label{app:data}

\begin{figure*}
\includegraphics[trim = {.25cm 4.7cm 3.5cm 2.75cm},clip,width=.9\textwidth]{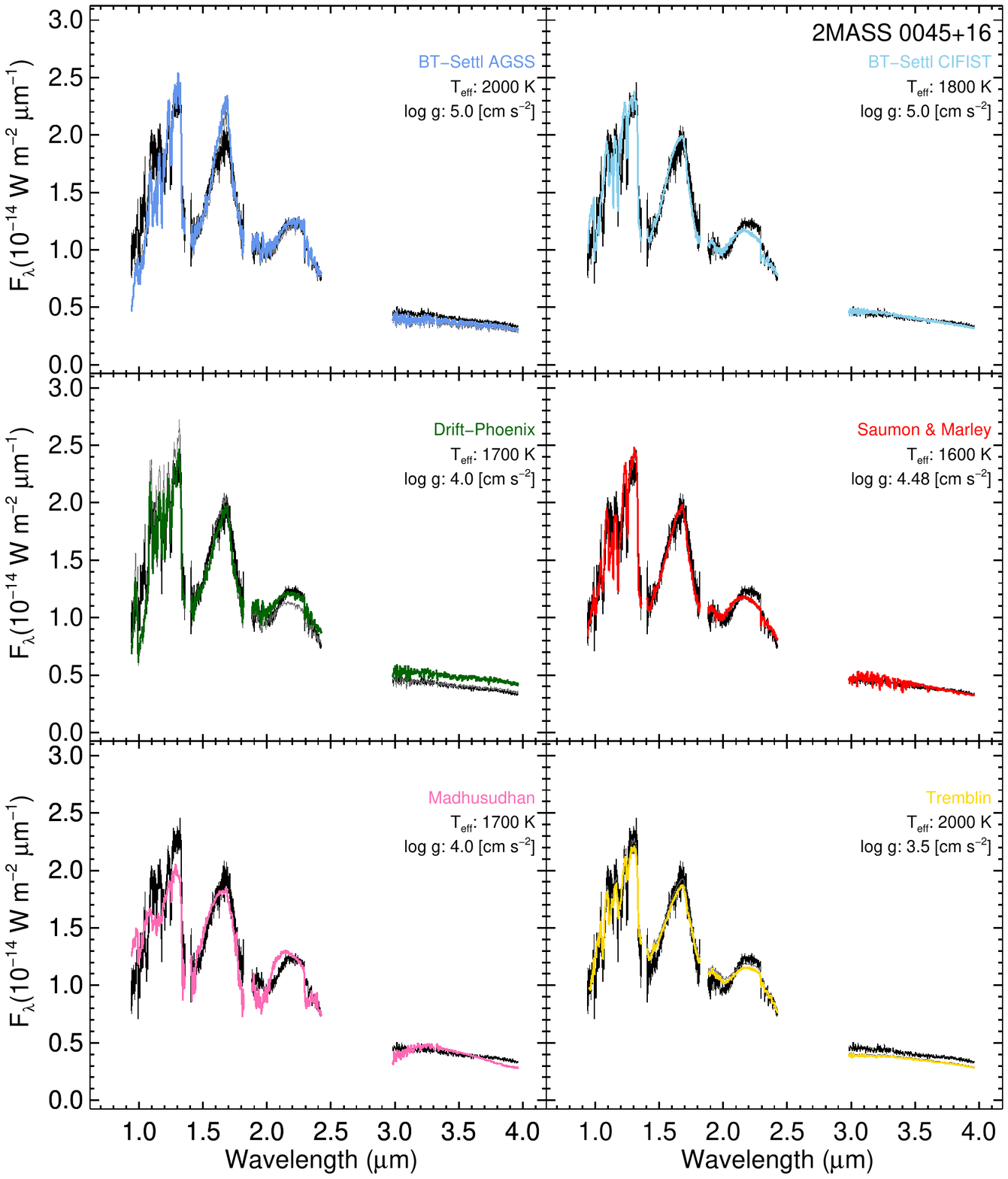}
\centering
\caption{The combined spectrum of 2MASS 0045+16 (black) compared to the model spectra (colored). For ease of comparison, the near-IR best-fit models are shown in grey.}\label{fig:0045Fits}
\end{figure*}

\begin{figure*}
\includegraphics[trim = {.25cm 4.7cm 3.5cm 2.75cm},clip,width=.9\textwidth]{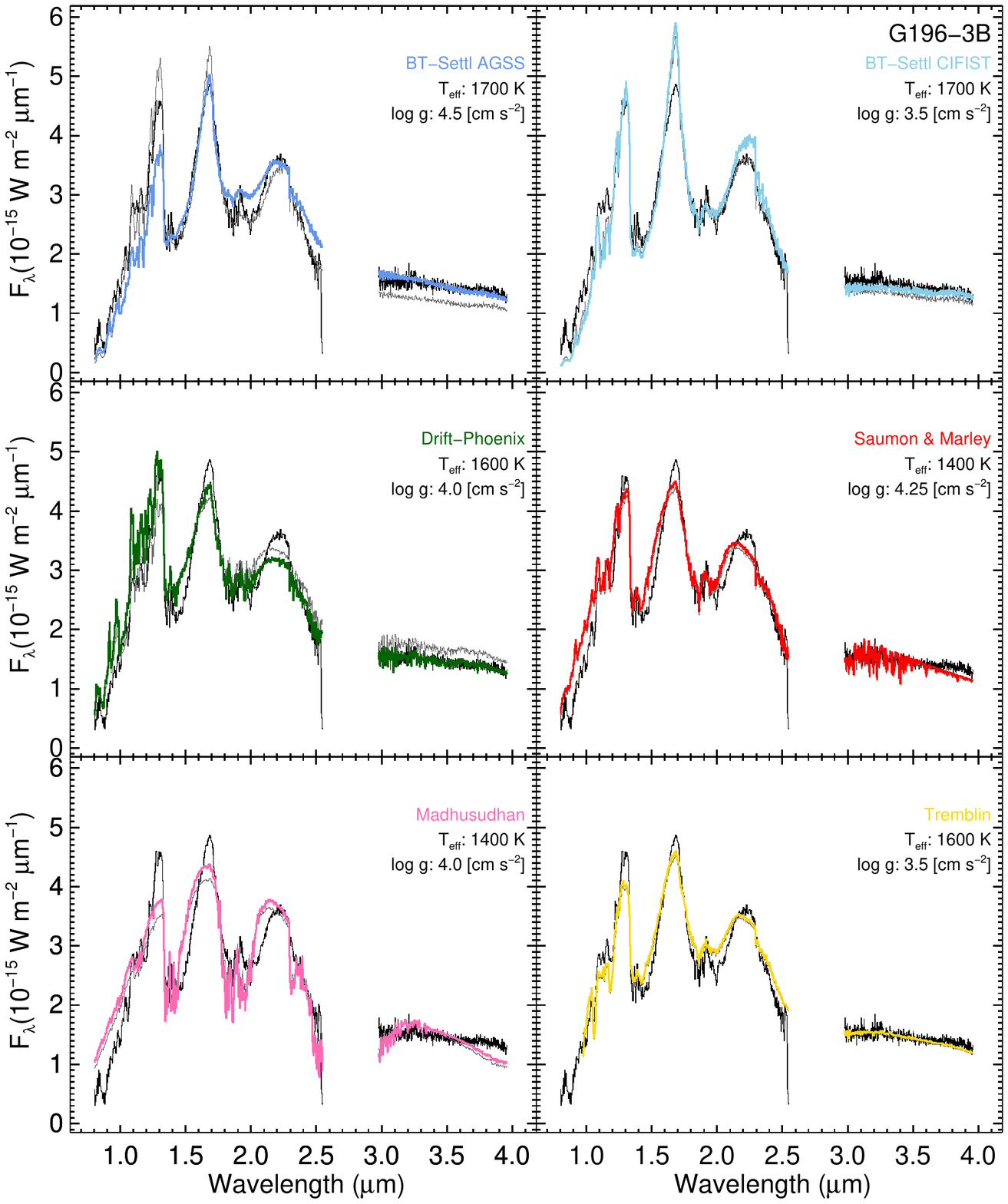}
\centering
\caption{The combined spectrum of G196-3B (black) compared to the model spectra (colored). For ease of comparison, the near-IR best-fit models are shown in grey.}\label{fig:G196Fits}
\end{figure*}

\begin{figure*}
\includegraphics[trim = {.25cm 4.7cm 3.5cm 2.75cm},clip,width=.9\textwidth]{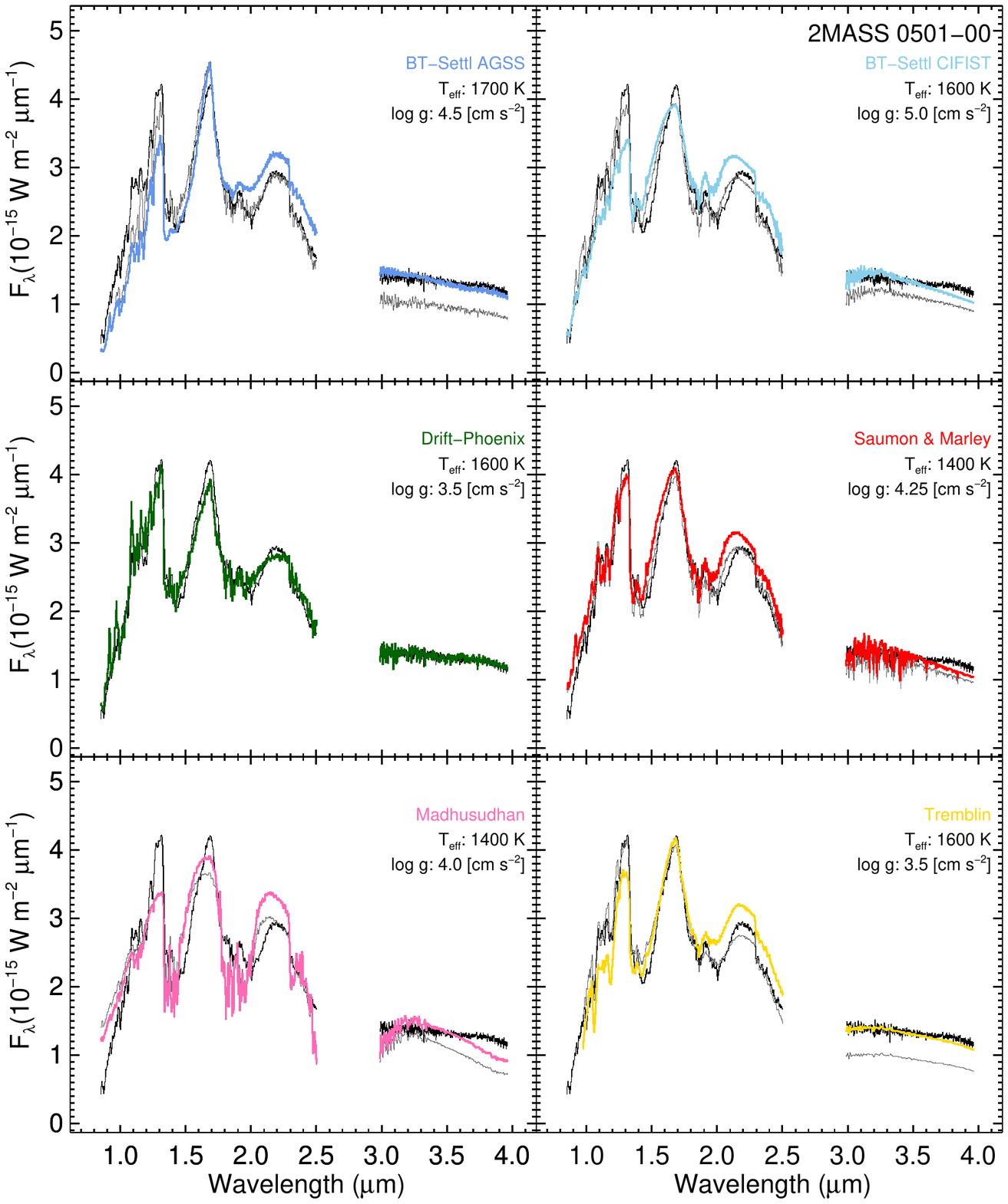}
\centering
\caption{The combined spectrum of 2MASS 0501-00 (black) compared to the model spectra (colored). For ease of comparison, the near-IR best-fit models are shown in grey.}\label{fig:0501Fits}
\end{figure*}

\begin{figure*}
\includegraphics[trim = {.25cm 4.7cm 3.5cm 2.75cm},clip,width=.9\textwidth]{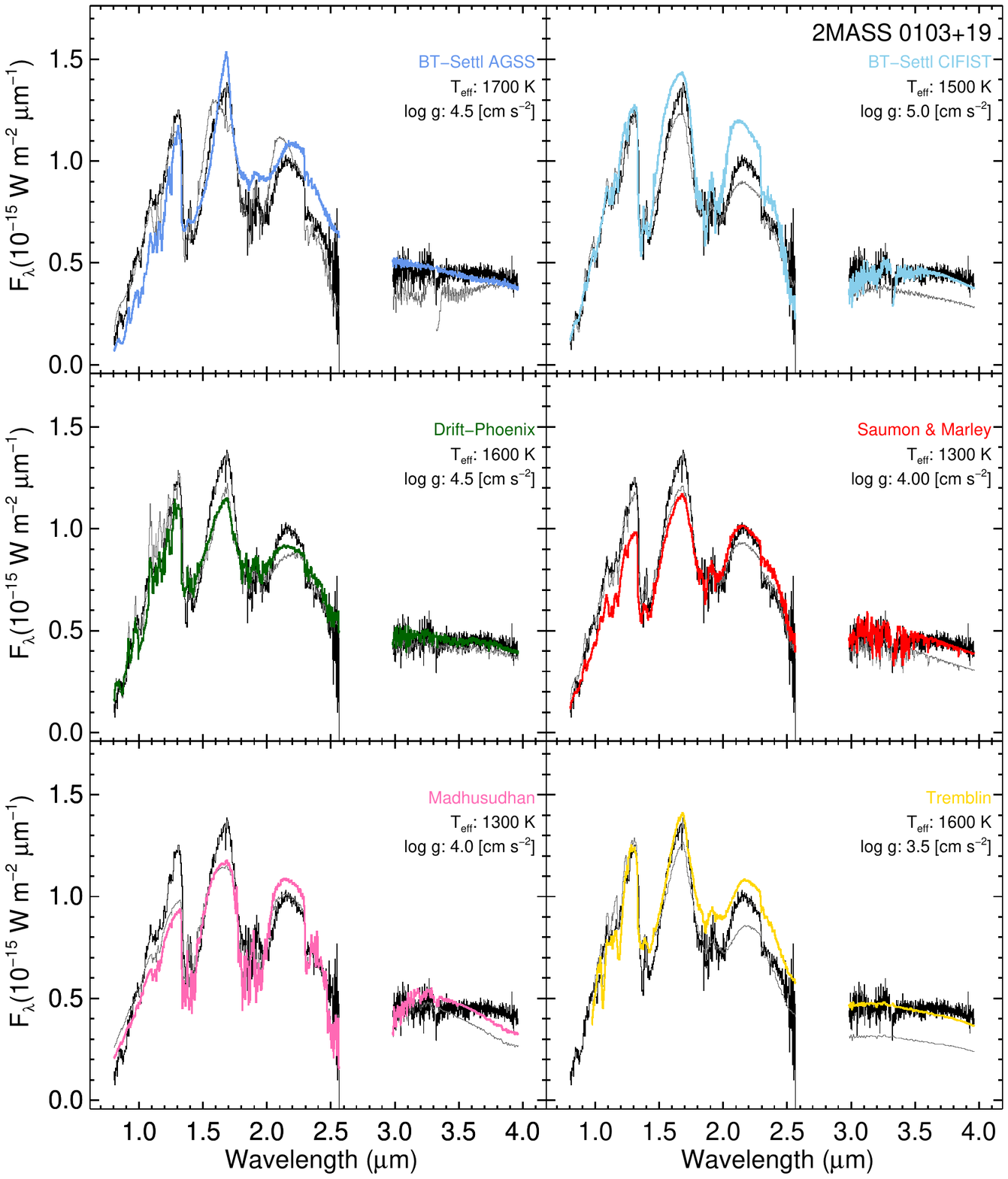}
\centering
\caption{The combined spectrum of 2MASS 0103+19 (black) compared to the model spectra (colored). For ease of comparison, the near-IR best-fit models are shown in grey.}\label{fig:0103Fits}
\end{figure*}

\begin{figure*}
\includegraphics[trim = {.25cm 4.7cm 3.5cm 2.75cm},clip,width=.9\textwidth]{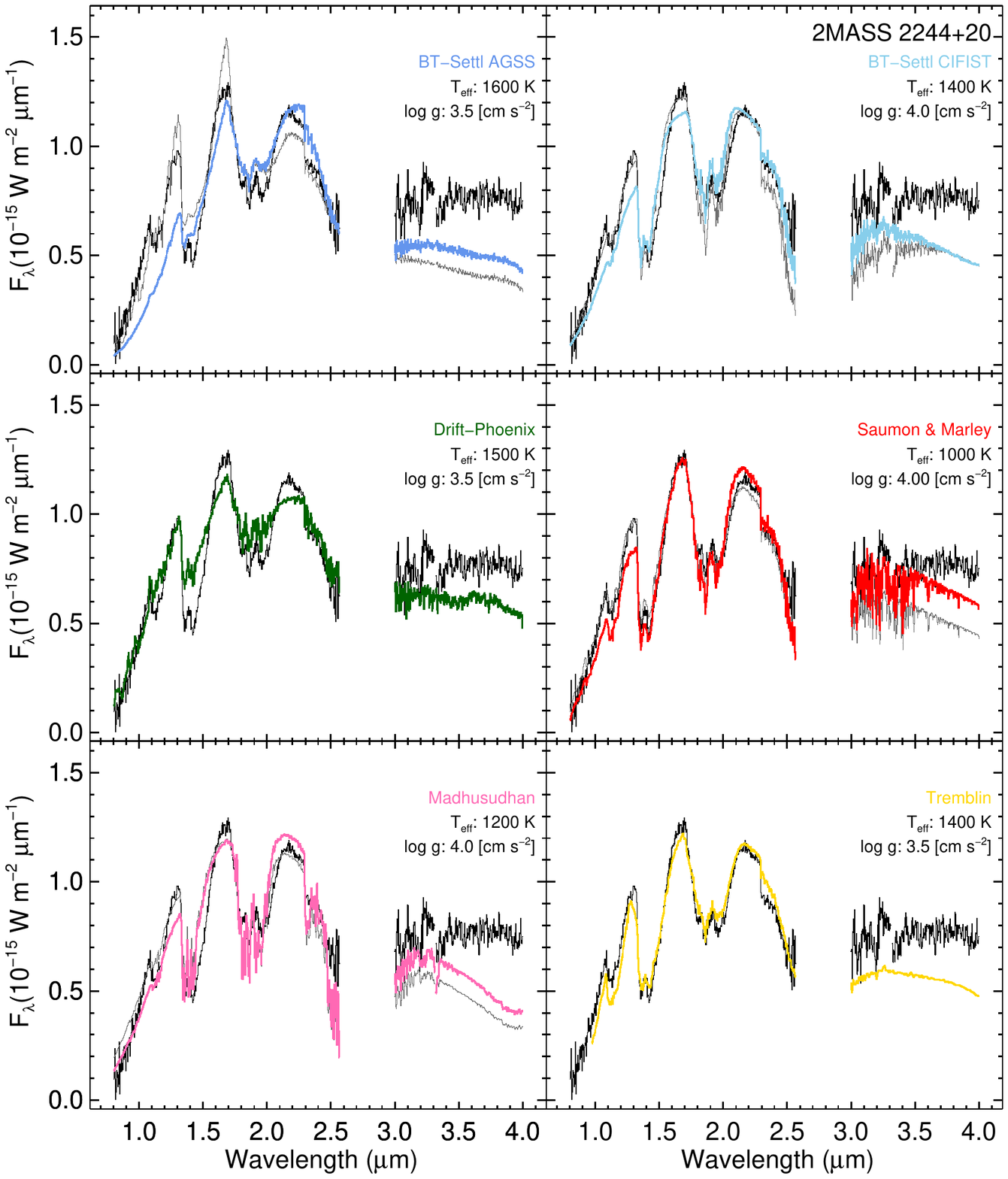}
\centering
\caption{The combined spectrum of 2MASS 2244+20 (black) compared to the model spectra (colored). For ease of comparison, the near-IR best-fit models are shown in grey.}\label{fig:2244Fits}
\end{figure*}

\begin{figure*}
\includegraphics[trim = {.25cm 4.7cm 3.5cm 2.75cm},clip,width=.9\textwidth]{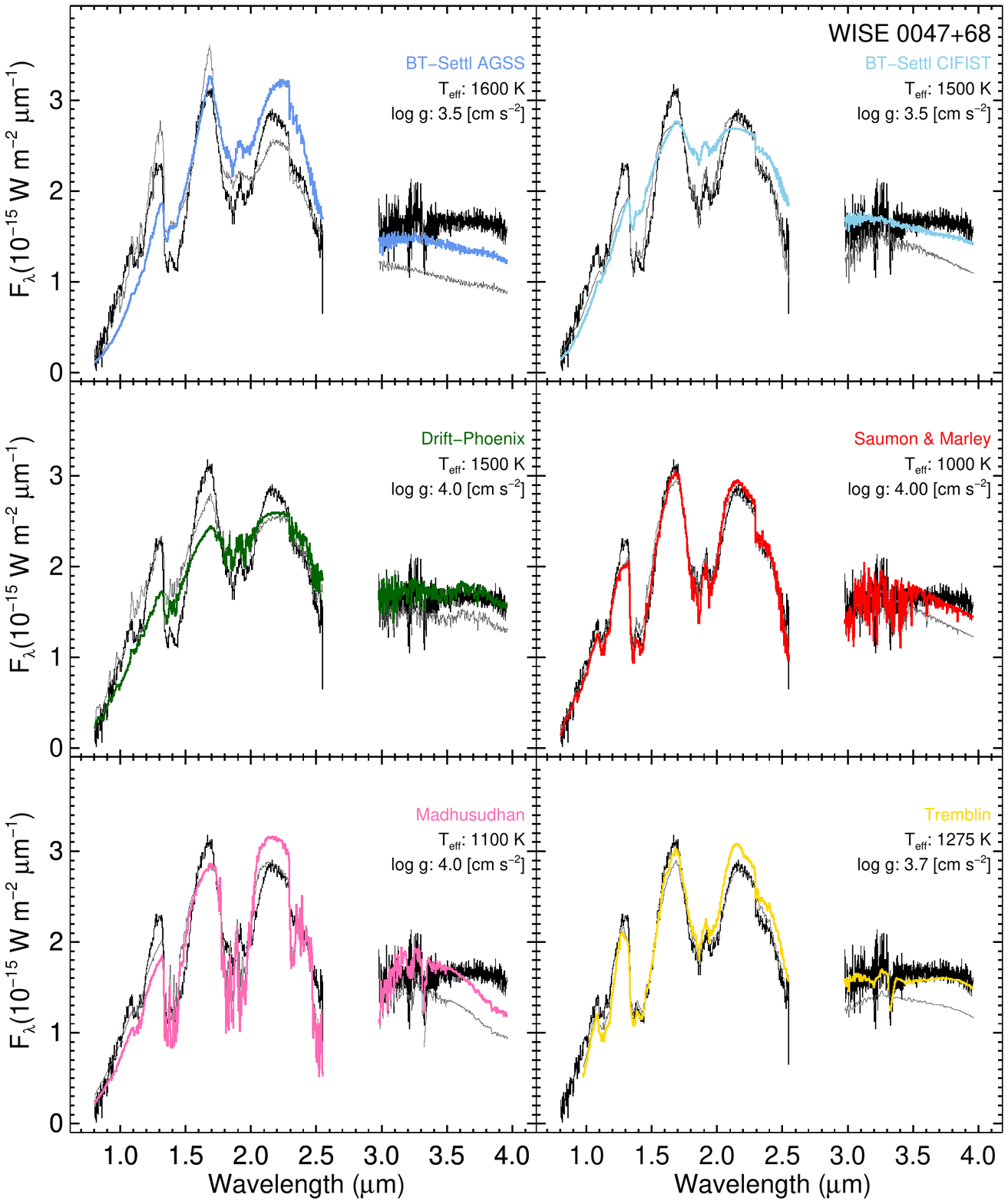}
\centering
\caption{The combined spectrum of WISE 0047+68 (black) compared to the model spectra (colored). For ease of comparison, the near-IR best-fit models are shown in grey.}\label{fig:0047Fits}
\end{figure*}

Figures \ref{fig:0045Fits} through \ref{fig:0047Fits} show the combined spectra and best-fit model for all the objects not already found in the paper (See Figure \ref{fig:0355Fits} and Figure \ref{fig:318Fits} for 2MASS 0355+11 and PSO 318.5-22).

\bsp	
\label{lastpage}
\end{document}